\documentclass[usenatbib]{mn2e}
\usepackage{psfig}

\title[Modelling of HST lightcurves]{Hubble Space Telescope
Observations of SV Cam: II. First Derivative Lightcurve Modelling using
{\sc phoenix} and {\sc atlas} Model Atmospheres}

\author[S.V. Jeffers et al.]
       {S.V.Jeffers$^{1,2}$, J.P.Aufdenberg$^{3}$, G.A.J.Hussain$^{4}$,
A.Collier Cameron$^2$, V.R.Holzwarth$^2$ \\
$^1$ Laboratoire d'Astrophysique de Toulouse-Tarbes, Observatoire
Midi-Pyr$\acute{e}$n$\acute{e}$es,14, avenue Edouard Belin, F-31400 Toulouse, France\\
        $^2$ School of Physics and Astronomy, University of St
Andrews, North Haugh, St Andrews,
Fife KY16 9SS, U.K.\\
$^3$ National Optical Astronomy Observatory, 950 N. Cherry Ave,
Tucson, AZ 85726, U.S.A. \\
$^4$ Astrophysics Div., Research \& Science Support Department of ESA,
ESTEC, Postbus 299, Noordwijk, The Netherlands \\ 
}

\date{}

\pagerange{\pageref{firstpage}--\pageref{lastpage}}
\pubyear{2005}

\begin{document}

\maketitle

\label{firstpage}

\begin{abstract}

The variation of the specific intensity across the stellar disc is
essential input parameter in surface brightness reconstruction
techniques such as Doppler imaging, where the relative intensity
contributions of different surface elements are important in detecting
starspots.  We use {\sc phoenix} and {\sc atlas} model atmospheres to
model lightcurves derived from high precision (S/N $\simeq$ 5000) HST
data of the eclipsing binary SV Cam (F9V + K4V), where the variation
of specific intensity across the stellar disc will determine the
contact points of the binary system lightcurve.  For the first time we
use $\chi^2$ comparison fits to the first derivative profiles to
determine the best-fitting model atmosphere.  We show the wavelength
dependence of the limb darkening and that the first derivative profile
is sensitive to the limb-darkening profile very close to the limb of
the primary star.  It is concluded that there is only a marginal
difference ($<$ 1$\sigma$) between the $\chi^2$ comparison fits of the
two model atmospheres to the HST lightcurve at all wavelengths.  The
usefulness of the second derivative of the light-curve for measuring
the sharpness of the primary's limb is investigated, but we find that
the data are too noisy to permit a quantitative analysis.

\end{abstract}


\section{Introduction}

Limb darkening effects in stellar atmospheres have important
implications throughout stellar astrophysics where a determination of
the surface brightness distribution is important.  Recent work using
Doppler imaging and micro-lensing events has shown that commonly-used
analytical limb darkening laws fail to match stellar observations at
the limb of the star \citep{thurl04,barnes04aephe}.  Other
micro-lensing results \citep{fields03} show that the intensity
predictions from model atmospheres are discrepant in the case of a
K-giant. 

Surface brightness reconstruction techniques such as Doppler imaging
and eclipse mapping rely on the information content of surface areas
with different distances from the rotation axis (see review by
\citet{camerondoppler01}).  To detect starspots Doppler imaging uses 
the relative intensity contributions, calculated from model
atmospheres, of the different surface elements.  To
reconstruct an accurate surface brightness distribution it is
essential to know how parameters, such as the limb darkening, can
alter the intensity values across the stellar disc.

High inclination eclipsing binary systems can be used as probes to
determine the variation of specific intensity at the stellar limb.  If
limb darkening showed a smooth transition in specific intensity at the
limb of the star, the contact points of eclipses would appear less
abrupt and slightly displaced in phase relative to models with limb
darkening laws derived from plane-parallel atmospheres, where the
cutoff is very sharp.  The sharp cutoff in plane parallel atmospheres
results from the optical depth of the rays being infinite at the limb.

In November 2001 we were awarded 9 orbits of HST/STIS time to
eclipse-map the inner face of the F9V primary of the totally eclipsing
binary SV Cam.  SV Cam (F9V\,+\,K4V) is a synchronously rotating RS
CVn binary with a period of 0.59d.  We obtained spectrophotometric
lightcurves of 3 primary eclipses with a signal-to-noise ratio of
5000.  The first analysis of these data, by \citet{jeffersem05},
determined the radii of the primary and secondary stars.  When the
resulting lightcurve was subtracted from the observed data, the
residual lightcurve showed strong peaks at phases of
contact. \citet{jeffersem05} then showed that these mismatches are
reduced significantly, but not eliminated, when a polar cap and a
reduction in the photospheric temperature, to synthesise high spot
coverage, are imposed on the image.  

As there is a significant temperature difference between the primary
and secondary stars, the secondary star acts as a dark occulting disc
as it eclipses the primary star.  The variation of brightness as a
function of phase reflects the degree of limb darkening on the primary
star.  In this paper we determine the best fitting model atmosphere by
fitting the models to the brightness variations as the secondary scans
the inner face of the primary star, using the first and second
derivatives of the HST in 10 wavelength bands.  We discuss the
implications these results have on results from Doppler Imaging.
 
\section{Model atmospheres}

In this paper, two well established stellar atmosphere codes are used;
the {\sc phoenix} model atmosphere code  \citep{hauschildt99} that uses spherical
atmospheres and the {\sc atlas} model atmosphere code
\citep{kurucz94cdrom} that uses plane
parallel atmospheres.  

\subsection{{\sc atlas}}

We use the plane-parallel ATLAS9 model atmospheres from the Kurucz
CD-ROMS \citep{kurucz94cdrom}. We integrate the intensity values over
the wavelength range of our observations 2900\AA\ to 5700\AA.  We use
temperature models from 3500\,K to 6500\,K, with 250\,K interval,
across 17 limb angles.  The limb angle $\mu$ is defined by
$\mu$=\,cos\,$\theta$, where $\theta$ is the angle between the line of
sight and the normal vector of the local surface element.  The
treatment of convection is based on the mixing length theory with
approximate overshoot (ref. \cite{castelli97}) with a mixing length to
scale height ratio of 1.25.  The variation of specific intensity,
i.e. at $\mu$=1,as a function of wavelength and limb angle is shown in
Figure~\ref{centintwav}.

\subsection{{\sc phoenix}}

The general input physics set-up of the {\sc phoenix} model atmosphere
code is discussed in \citet{hauschildt99}.  The main advantage of
using this code is that it is based on spherical geometry (spherical
radiative transfer) LTE rather than traditional plane-parallel
structure.  NLTE effects are considered to be insignificant in this
application.

The synthetic spectra are based on an extension of the grid of PHOENIX
model atmospheres described by \citet{allard00}.  This extended grid
includes surface gravities larger than $\log(g) > 3.0$ needed for main
sequence stars.  These models are as described by
\citet{hauschildt99}, but include an updated molecular line list.  The
models are computed in spherical geometry with full atomic and
molecular line blanketing using solar elemental abundances.  In these
models, the stellar mass is 0.5 M$_\odot$ and the convection treatment
assumes a mixing-length to pressure scale height ratio of 2 with no
overshooting.  There are 117 synthetic spectra in total.  The
effective temperature runs from 2700K to 6500 K in 100K steps at three
surface gravities: $\log(g)$ = 4.0, 4.5, and 5.0.  The wavelength
resolution of these synthetic spectra is 1\AA.

\begin{figure}
\hspace{0.25cm}
\psfig{file=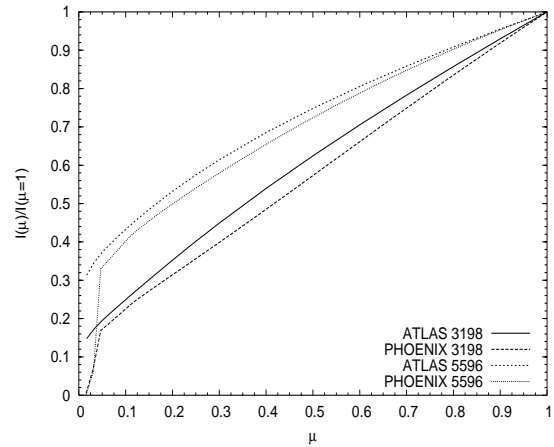,width=7.5cm,height=6cm,angle=270}
\caption{The predicted variation in specific intensity across the stellar disc as a 
function of wavelength for the longest (5596\AA) and the shortest (3198\AA) wavelengths in
this analysis (T=6000\,K, log g=4.5).}
\protect\label{centintwav}
\end{figure}

The variation of specific intensity as a function of limb angle is
shown in Figure~\ref{centintwav}.  The difference between the
{\sc phoenix} and the {\sc atlas} model atmospheres at the limb of the
star results primarily from the effects of spherical geometry of the
{\sc phoenix} model.  In models using spherical geometry, there is a
finite optical depth for rays close to the limb, while with plane
parallel models the optical depth of the rays is infinite at the limb
providing significant intensities down to $\mu$=0.  Other contributing
effects include overshooting and the mixing length ratio.

In addition to the limb effects, Fig.~\ref{centintwav} shows that {\sc
atlas} intensity profiles are overall brighter than {\sc phoenix}
intensity profiles.  We attribute this difference to the use of
overshooting in the {\sc atlas} models.  {\sc atlas} models with
overshooting have been shown to have weaker limb darkening in the blue
relative to models without overshooting, for instance, overshooting
models have been shown to better fit solar limb darkening observations
than non-overshooting models \cite{castelli97}.  A similar result
holds for Procyon (F5 IV-V); ref. \cite{aufdenberg05} for a detailed
study of the effects of 1-D and 3-D convection on limb darkening.
  
\subsection{Model lightcurves}

We use the eclipse-mapping code DoTS \citep{cameron97dots} to model
the primary eclipse lightcurve for each model atmosphere, and to
determine which model atmosphere best fits the complete HST
lightcurve. The input data to each model comprises: the variation of
the specific intensity as a function of limb angle, as shown in
Figure~\ref{centintwav} for both model atmospheres considered; the
primary and secondary radii; a reduced primary photospheric
temperature and polar spot.  Gravity darkening is also included 
according to the description of \cite{cameron97dots}.

We include a reduced photospheric temperature to account for a stellar
surface that is peppered with small spots which are too small to be
resolved through eclipse mapping.  Following the results of
\citet{jeffersem05} we include a polar cap in the modelled lightcurve
to optimise the fit of the lightcurve to the data.  The method for
modelling the photometric lightcurve to include a reduced photospheric
temperature and a polar cap is described in Appendix A for the {\sc
atlas} model atmosphere.  The binary system parameters are summarised
in Table~\ref{t-param}.  The lightcurve solutions for the two model
atmospheres are in good agreement.

\section{HST Observations}

\begin{table}


\begin{tabular}{l c c c c c l }

\hline
\hline

{Visit}& {Obs. Date} & {UT} & {UT} & {Exposure} & {No of} \\ & &
{Start} & {End} & {Time(s)} & {Frames}& \\

\hline

1 & {01 November 2001} & 20:55:56 & 01:00:17 & 30 & 165 \\
2 & {03 November 2001} & 14:34:29 & 18:38:21 & 30 & 165 \\
3 & {05 November 2001} & 09:49:17 & 13:52:22 & 30 & 165 \\

\hline
\hline
\end{tabular}
\caption{HST Observations of SV Cam, where the exposure time is 
per frame.}
\protect\label{obsl}
\end{table}

\begin{figure}
\protect\label{t-param}
\psfig{file=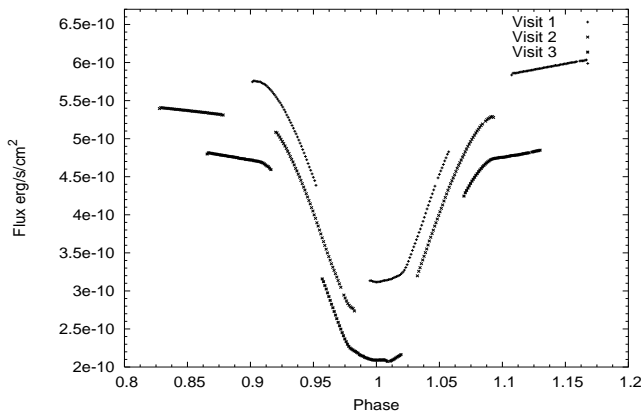,width=8.7cm,height=5.6cm,angle=270}
\caption{The comprising sections of the primary eclipse from each of
the 3 HST visits.  The flux density has been summed between
2900\,\AA\, and 5700\,\AA.  An off-set of 4$\times$10$^6$
ergs/s/cm$^2$ has been added for clarity.}
\protect\label{visit}
\end{figure}

Three primary eclipses of SV Cam were observed by the HST, using the
Space Telescope Imaging Spectrograph with the G430L grating.  The
observations used 9 spacecraft orbits and spanned 5 days at 2 day
intervals from 1-5 November 2001 as shown in Table 1.  Summing
the recorded counts over the observed wavelength range 2900\,\AA\, to
5700\,\AA\ yields a photometric lightcurve.  Figure~\ref{visit}
shows the photometry of the 3 eclipses observed during the 9 orbits.
The observations have a cadence of 40s and a photometric precision of
0.0002 magnitudes (S:N 5000) per 30\,s exposure. The observations and
the data reduction method are explained in greater detail in
\citet{jeffersem05}.

\begin{table*}
\fontsize{8}{12}\selectfont  
\begin{tabular}{l c c c c c c c}


\hline
\hline

Model Grid & Log g & L/H & Fitted & Spotted  & Primary & Secondary & Polar Spot \\

 & & & Temp (K) & Temp. (K) & Radius (r$_\odot$) & 
Radius (r$_\odot$) & (degrees) \\

\hline

{\sc phoenix} & 4.5 & 2.0 & 6038$\pm$58 & 5935\,$\pm$28 K & 1.235 $\pm$ 0.003 & 0.727  $\pm$
0.003 & 46.5 $\pm$ 8 \\
{\sc atlas} & 4.5 & 1.25 & 5972$\pm$59  & 5840\,$\pm$53 K & 1.241 $\pm$ 0.003 & 0.729 $\pm$ 0.002
& 45.7 $\pm$ 9 \\

\hline
\hline
\end{tabular}
\caption{Geometric binary system parameters computed for SV Cam
with a reduced photospheric temperature due to the star being peppered
with small spots, and increased radii due to the presence of a polar
cap.  L/H is the mixing-length to pressure scale height ratio}
\end{table*}

\section{Wavelength dependence of limb darkening}

\begin{figure}
\psfig{file=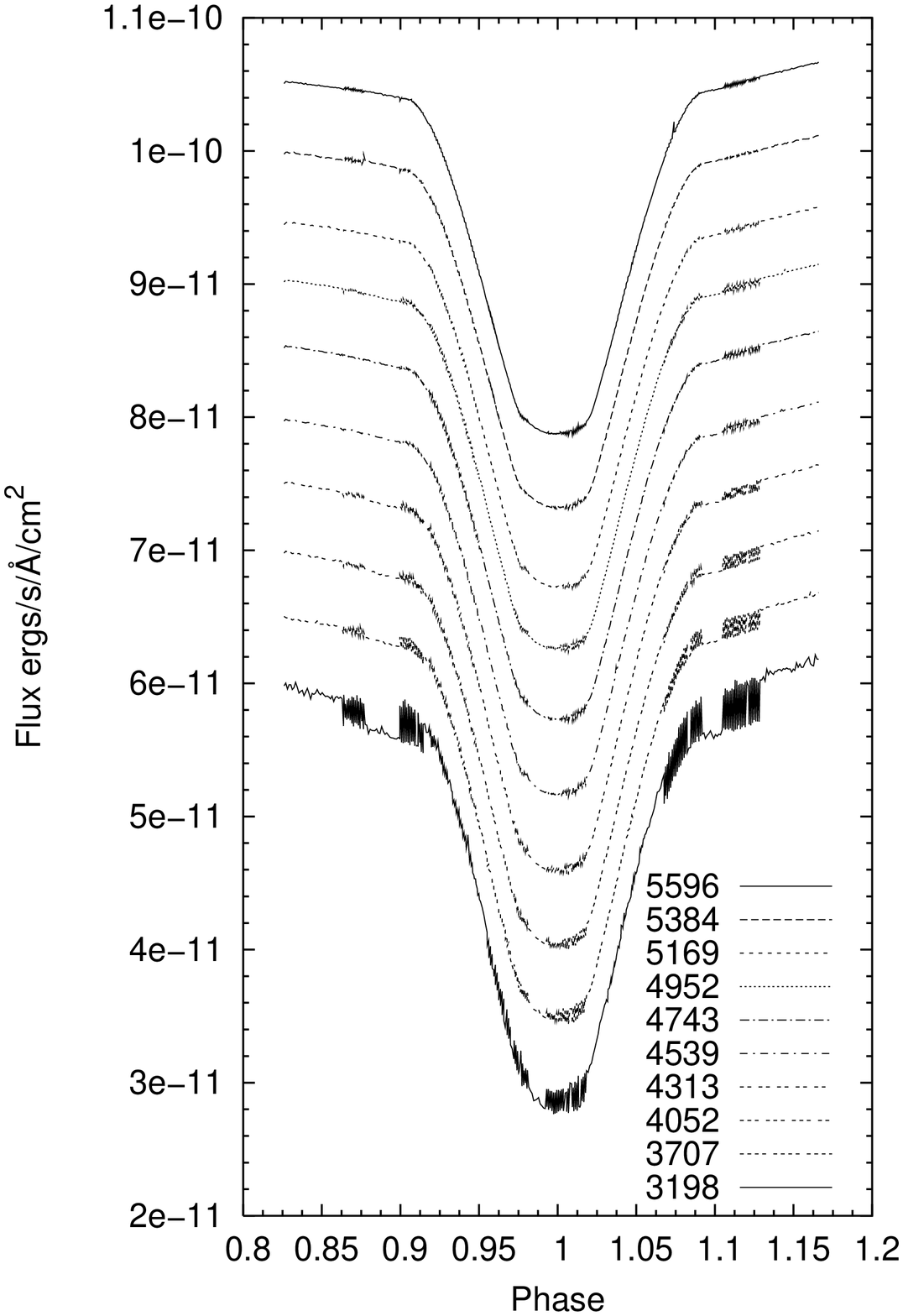,width=8.5cm,height=15cm,angle=0}
\caption{The variation of the HST lightcurve with wavelength.  Each
lightcurve has been plotted with an offset of 5 $\times$ 10$^{-12}$
from the lightcurve at the bottom of the plot (3198 \AA).}
\protect\label{1to10}
\end{figure}

\begin{figure}
\hspace{0.25cm}
\psfig{file=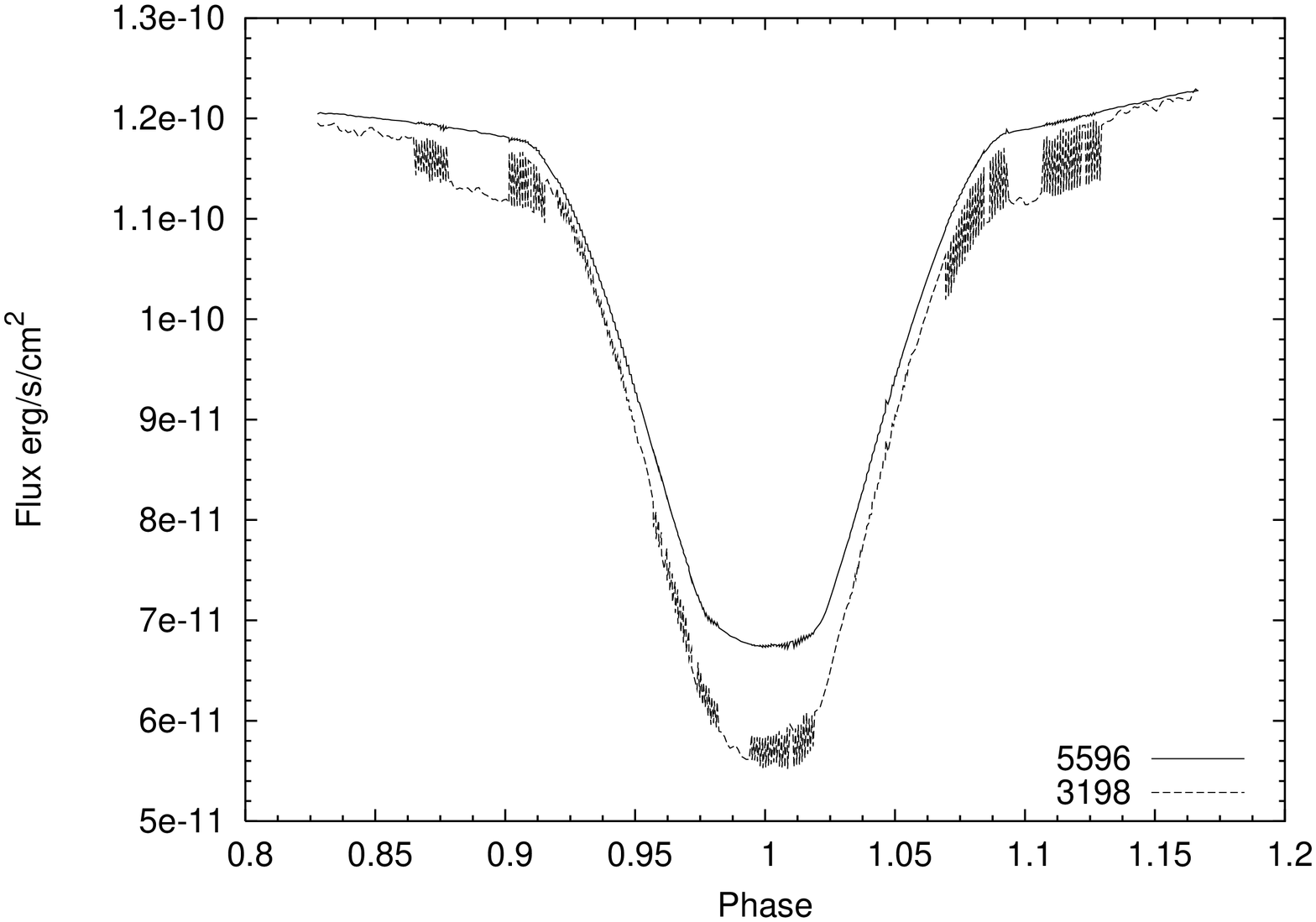,width=7.5cm,height=6cm,angle=270}
\caption{HST lightcurves with the longest (5596\,\AA) and shortest
wavelengths (3198\,\AA), showing the variation of limb darkening with
wavelength and the temporal evolution of bright magnetic features over
the 5 day timespan our observations.  The variation in the 3198\AA\
band is of the order of 5\%.}
\protect\label{1n10}
\end{figure}


\begin{figure}
\psfig{file=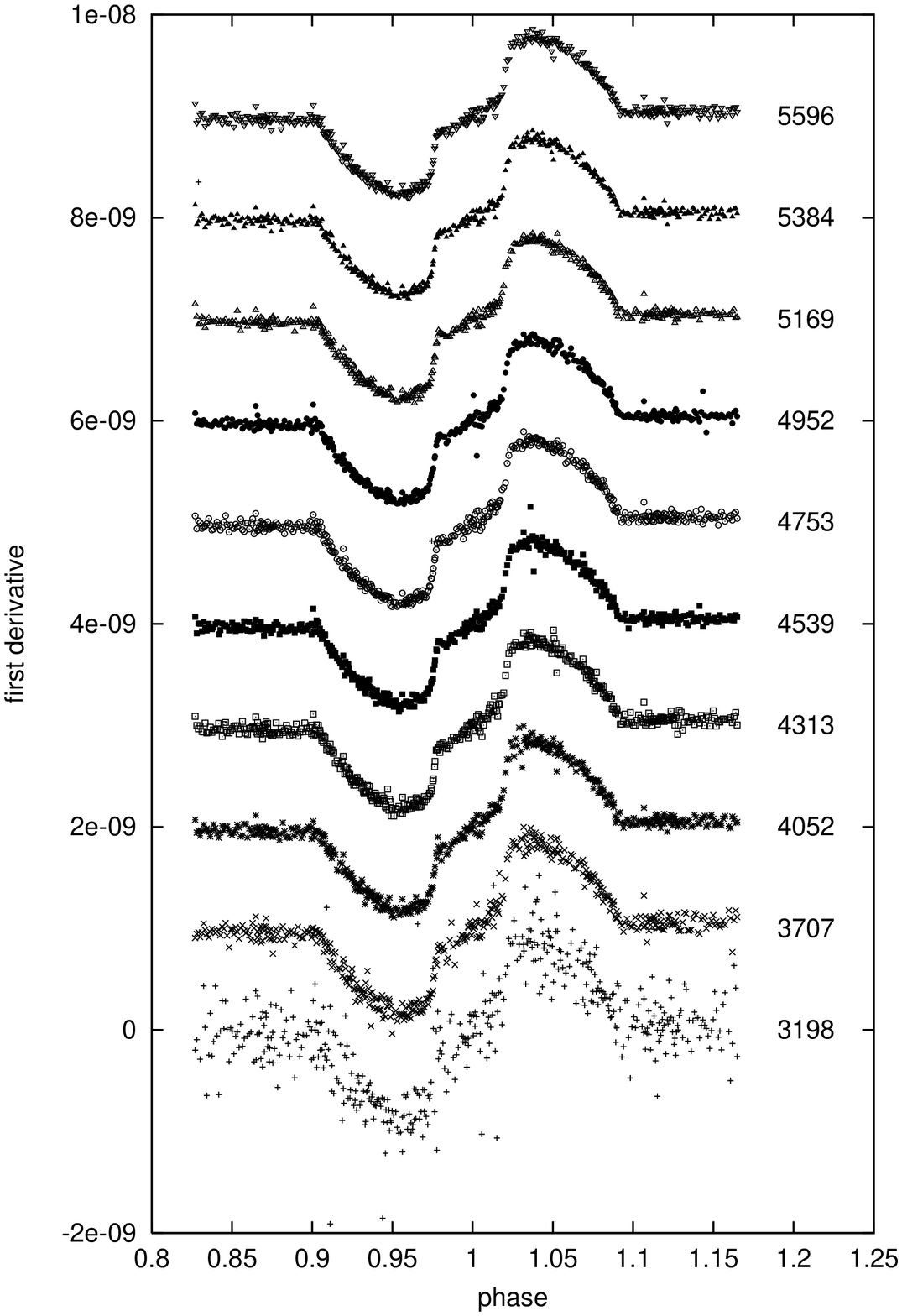,width=8cm,height=15cm,angle=0}
\caption{First derivative profiles for each of the 10 HST lightcurves 
as plotted in Figure~\ref{1to10}.  The presence of a bright magnetic
feature is the cause of the variation in waveband 3198\AA.}
\protect\label{fd1to10}
\end{figure}

\begin{figure}
\psfig{file=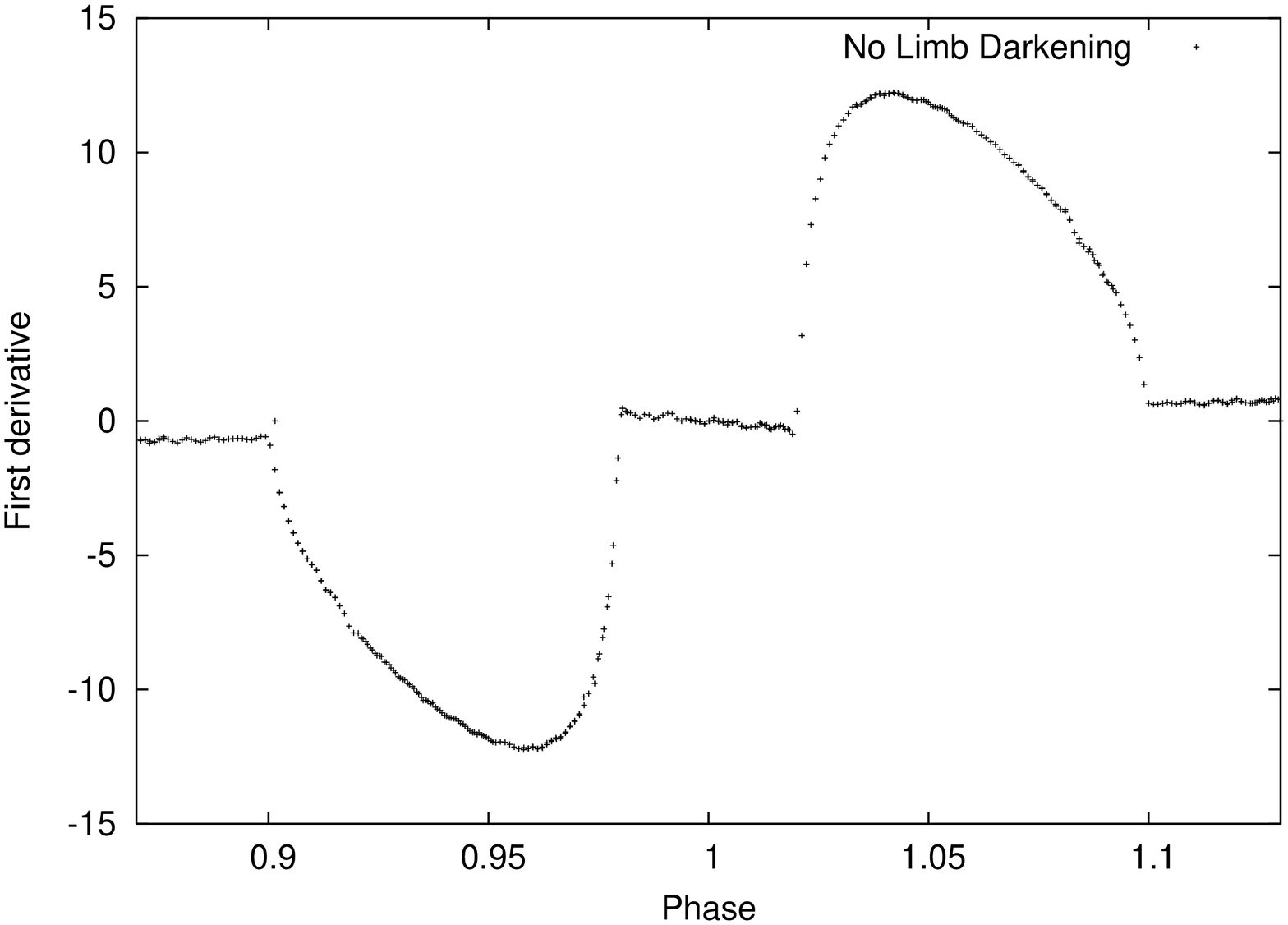,width=7.5cm,height=6cm,angle=270}
\caption{First derivative profiles, constructed with the same stellar
parameters as in Figure~\ref{fd1to10}, but with no limb darkening.
The slight negative slope during total eclipse results from a 
non-spherical star caused by tidal interactions in the binary system.}
\protect\label{fnld}
\end{figure}

\begin{figure}
\psfig{file=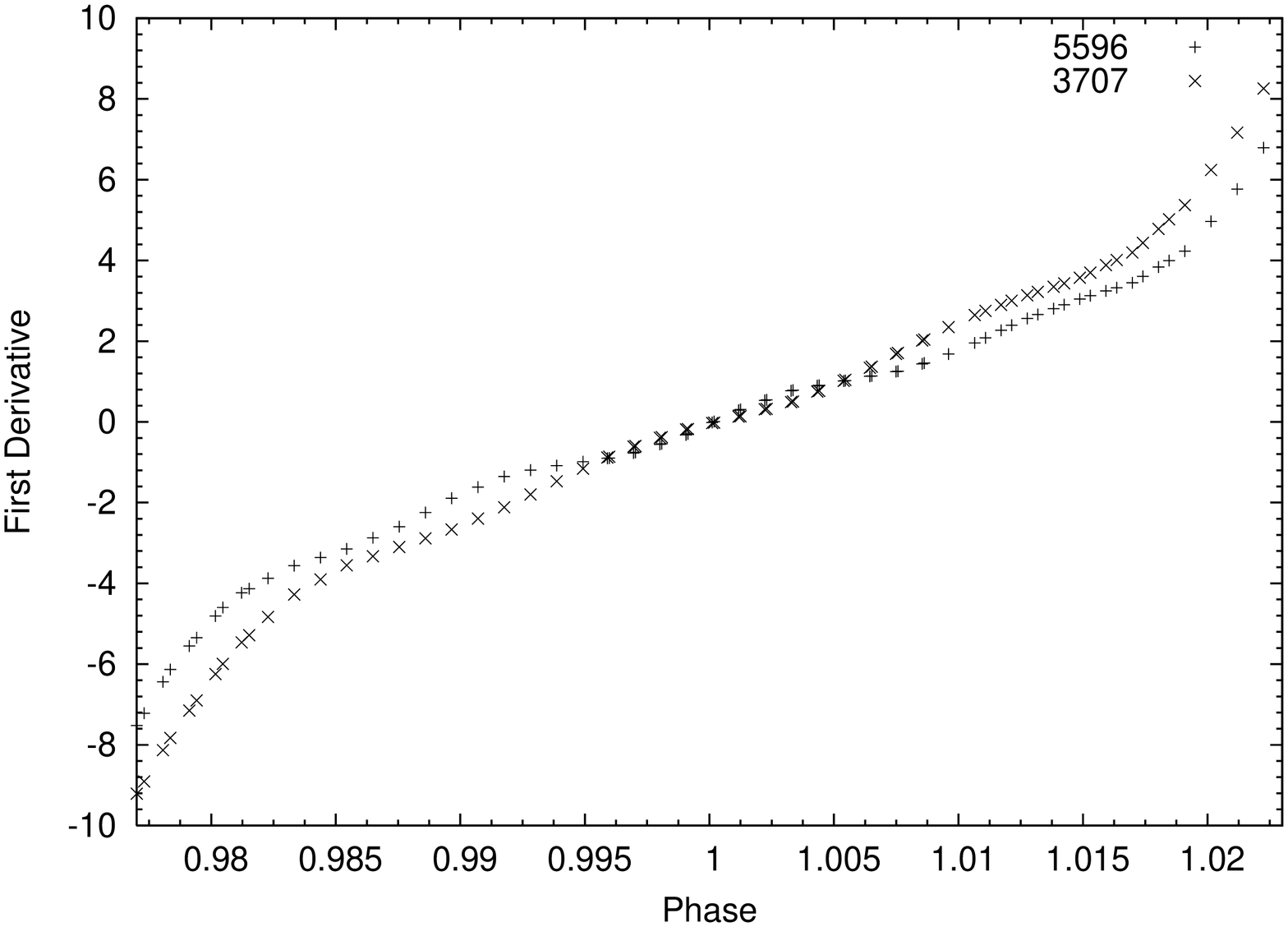,width=7.5cm,height=6cm,angle=270}
\caption{First derivative profiles of the observed lightcurve of 
wavelengths 5596\,\AA\ and 3707\,\AA\, for total eclipse, i.e. between
second and third contact, showing the variation of limb darkening with
wavelength.}
\protect\label{fd2n10}
\end{figure}

\begin{figure}
\psfig{file=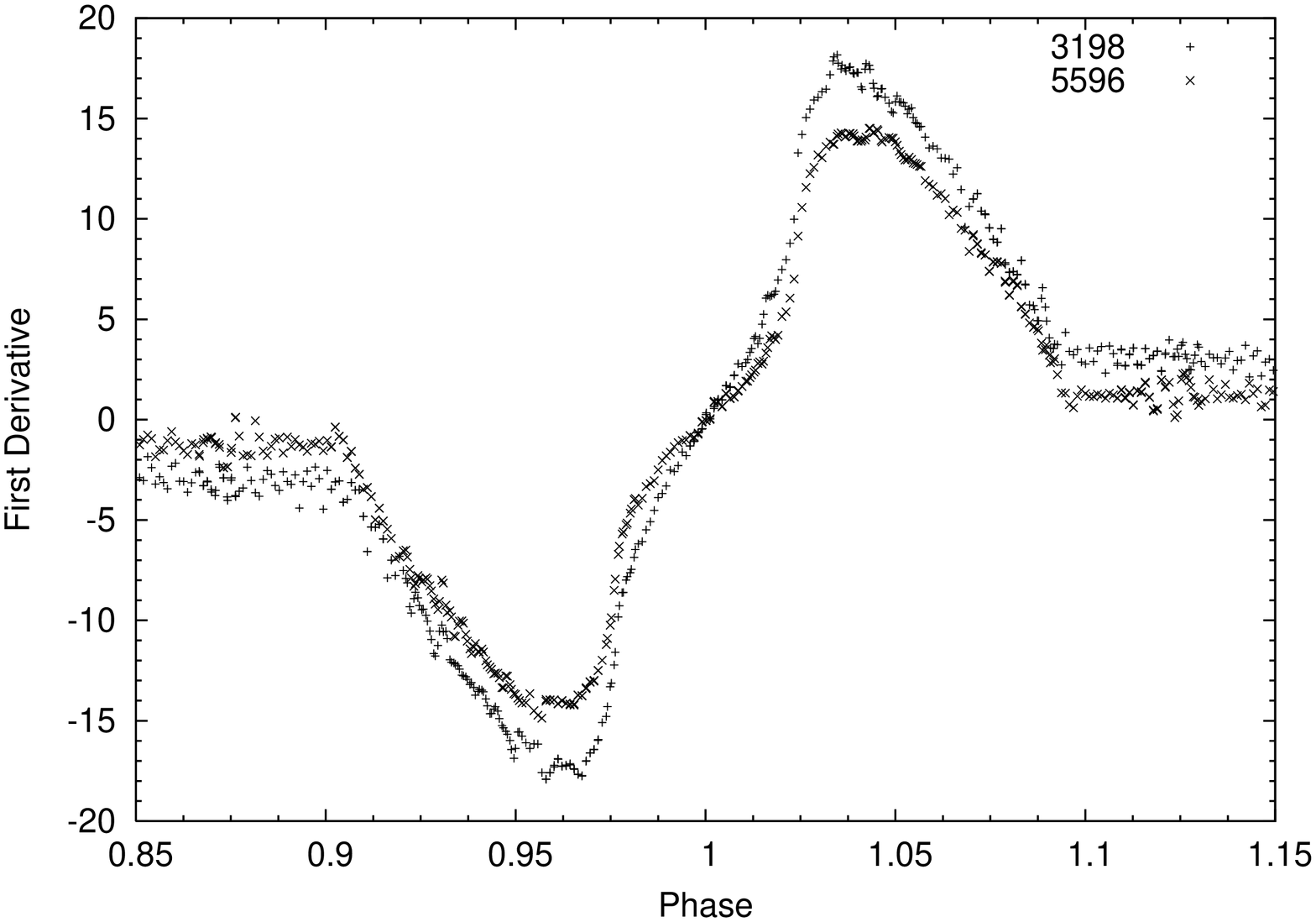,width=7.5cm,height=6cm,angle=270}
\caption{First derivative profiles for  the longest (5596\,\AA)
and shortest wavelengths (3198\,\AA), as plotted in
Figure~\ref{fd1to10}, using intensities interpolated from the {\sc
phoenix} grid at the best fit effective temperature and gravity 
shown in Table 2.}
\protect\label{phx-fd1n10}
\end{figure}

\begin{figure}
\psfig{file=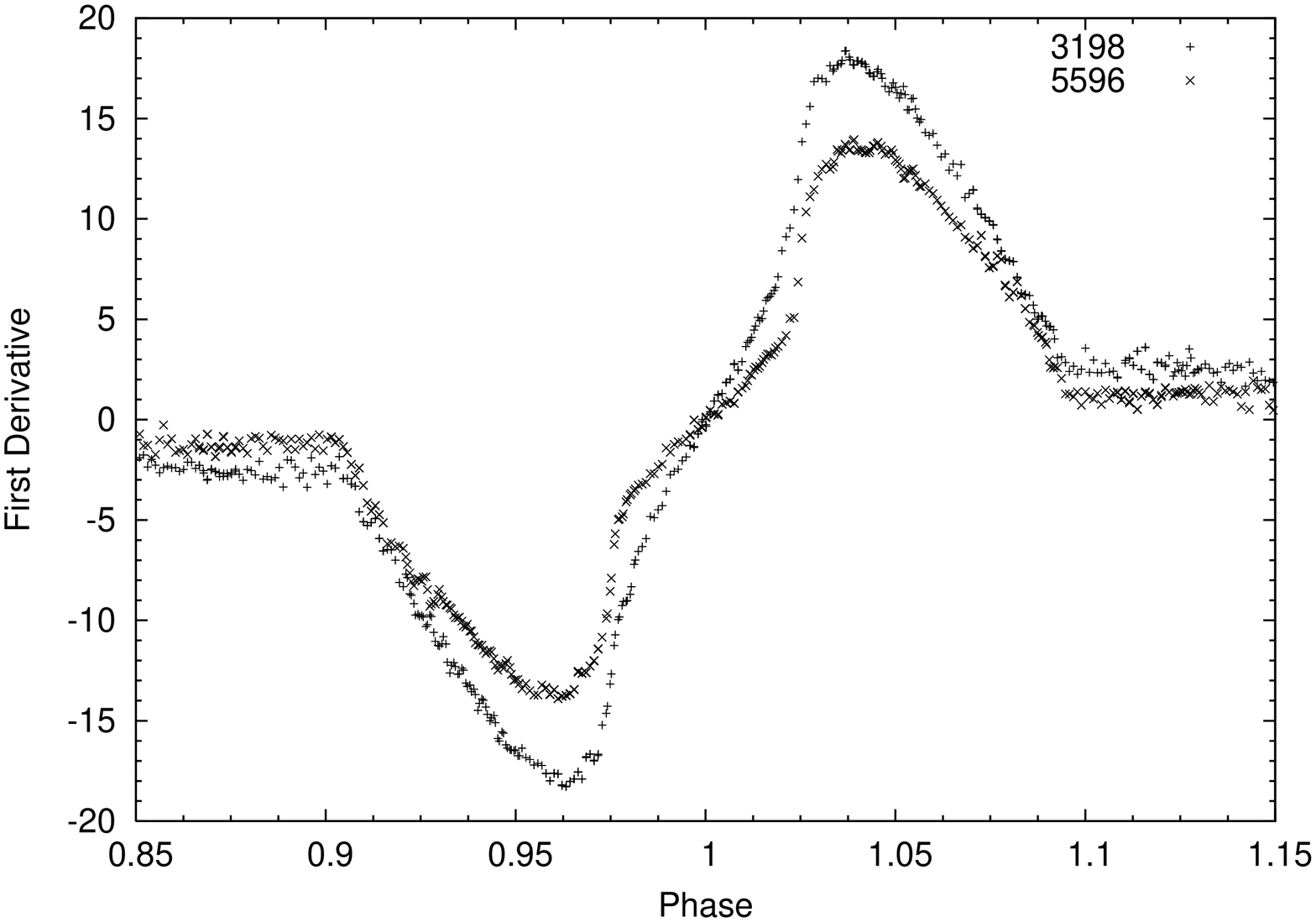,width=7.5cm,height=6cm,angle=270}
\caption{First derivative profiles for with the longest (5596\,\AA)
and shortest wavelengths (3198\,\AA), using intensities interpolated
from the {\sc atlas} grid at the best fit effective temperature and
gravity shown in Table 2.}
\protect\label{pp-fd1n10}
\end{figure}

The variation of the HST lightcurve with wavelength was determined by
phasing the three primary eclipses together and dividing each spectrum
into 10 bands of equal flux, as shown in Figure ~\ref{1to10}.  The
increased fluctuations at shorter wavelengths are consistent with the
presence of signatures bright magnetic activity, as they are strongest
in the bluest wavelength band.  The fluctuations at blue wavelengths
are not a short-timescale phenomenon, but rather a mismatch in flux
levels arising from a change in the star's UV flux from second to
third visit.

The curvature of the eclipse profile between second and third contact
illustrates clearly that limb darkening increases to wards shorter
wavelengths.  The wavelength dependence of limb-darkening is revealed
by the models collected by \citet{claret00ldc4}, but more
fundamentally by direct observations of the Sun, for example 
\citet{neckel94}.  Both the brightness variation and the
variation of the limb darkening with wavelength are illustrated by
plotting the bluest and reddest wavelength bands, as shown in
Figure~\ref{1n10}.

\section{First Derivative Profiles of Lightcurves}

\subsection{Basic properties}

As the primary star dominates the light of the binary system, the
cooler secondary acts as a dark occulting disc scanning across the
equatorial region of the primary during eclipse.  The variation
of the specific intensity as a function of limb angle across the primary
star shows the degree of limb darkening.  

To determine the optimally fitting model atmosphere to the HST
lightcurves it is necessary to examine the degree of curvature in the
eclipse profile.  We use the first derivative, with respect to phase,
of the lightcurves to determine the rate of change of the eclipsed
flux with phase.  This is the first time that this method has been
applied to photometric data of an eclipsing binary system.  The
numerical derivative is computed using the Interactive Data Language
(IDL) routines {\tt deriv} and {\tt derivsig} which employ 3-point
Lagrangian interpolation.

The profiles of the first derivative for each of the
10 HST wavebands are shown in Figure~\ref{fd1to10}.  The first contact
point is at phase 0.906$\pm$0.003, the second contact point is at
phase 0.977$\pm$0.003, the third contact point is at 1.023$\pm$0.003,
and the fourth contact point is at 1.093$\pm$0.003.  The gradient of
the first derivative profile between second and third contact points
is indicative of the degree of limb darkening in the lightcurve.  In
the first derivative of the model with no limb darkening, this region
the profile is flat (Figure~\ref{fnld}).  The variation of limb
darkening with wavelength is illustrated in Figure~\ref{fd2n10}, where
the first derivative lightcurves from the longest wavelength
(5596\,\AA) and the second shortest wavelength (3707\,\AA) are plotted
between the second and third contact points.  The first derivative
profile of the shortest wavelength contains too much deviation, caused
by a bright feature on the primary's surface, to provide a clear
example.

The variation of the first derivative profile with wavelength is also
visible in the profile using the {\sc phoenix} model atmosphere, as
shown in Figure~\ref{phx-fd1n10}.  In contrast to this
Figure~\ref{pp-fd1n10} shows the same first derivative profiles based
on the {\sc atlas} model atmosphere.  In the {\sc atlas} first
derivative profiles there is more variation between the two wavelength
bands than in the {\sc phoenix} model atmosphere.  The variation with
wavelength relates directly to the variation of specific intensity
with wavelength as previously shown in Figure~\ref{centintwav}.

\subsection{First derivative lightcurve fitting}

The first derivative profiles of the HST observations, in each of the
10 wavelength bands (Figure~\ref{fd1to10}), are fitted using reduced
$\chi^2$ comparison fits to the first derivatives of the {\sc phoenix}
and {\sc atlas} model atmospheres of the same wavelength as the
observations.  In this analysis the larger error bars are those of the
models, given the large uncertainties shown in Table 2, and the
finite number of planer elements used in the modelling of the
lightcurves.  We estimate these errors to be of the order of 1\%.  

\begin{figure}
\psfig{file=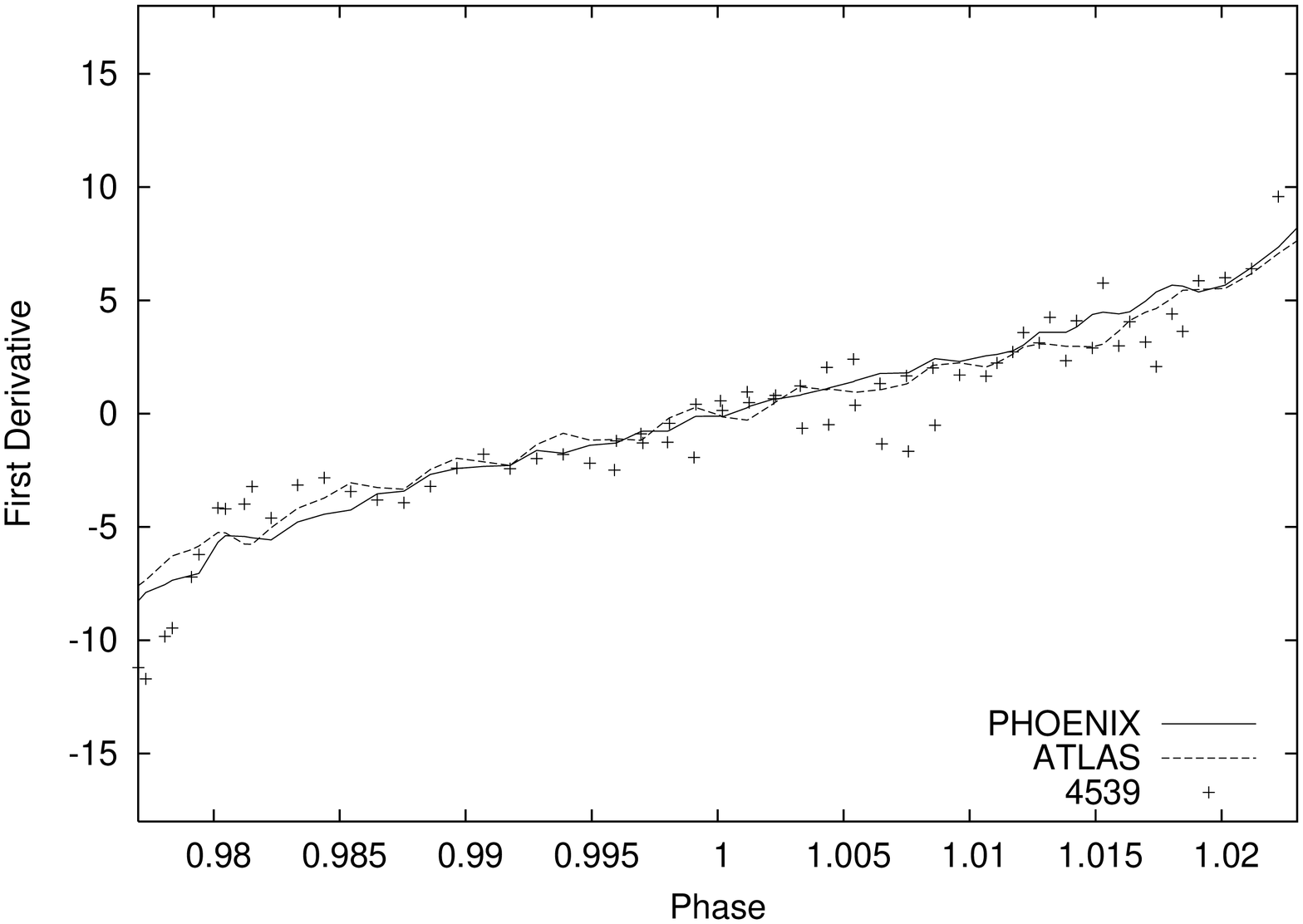,width=7.5cm,height=6cm,angle=270}
\psfig{file=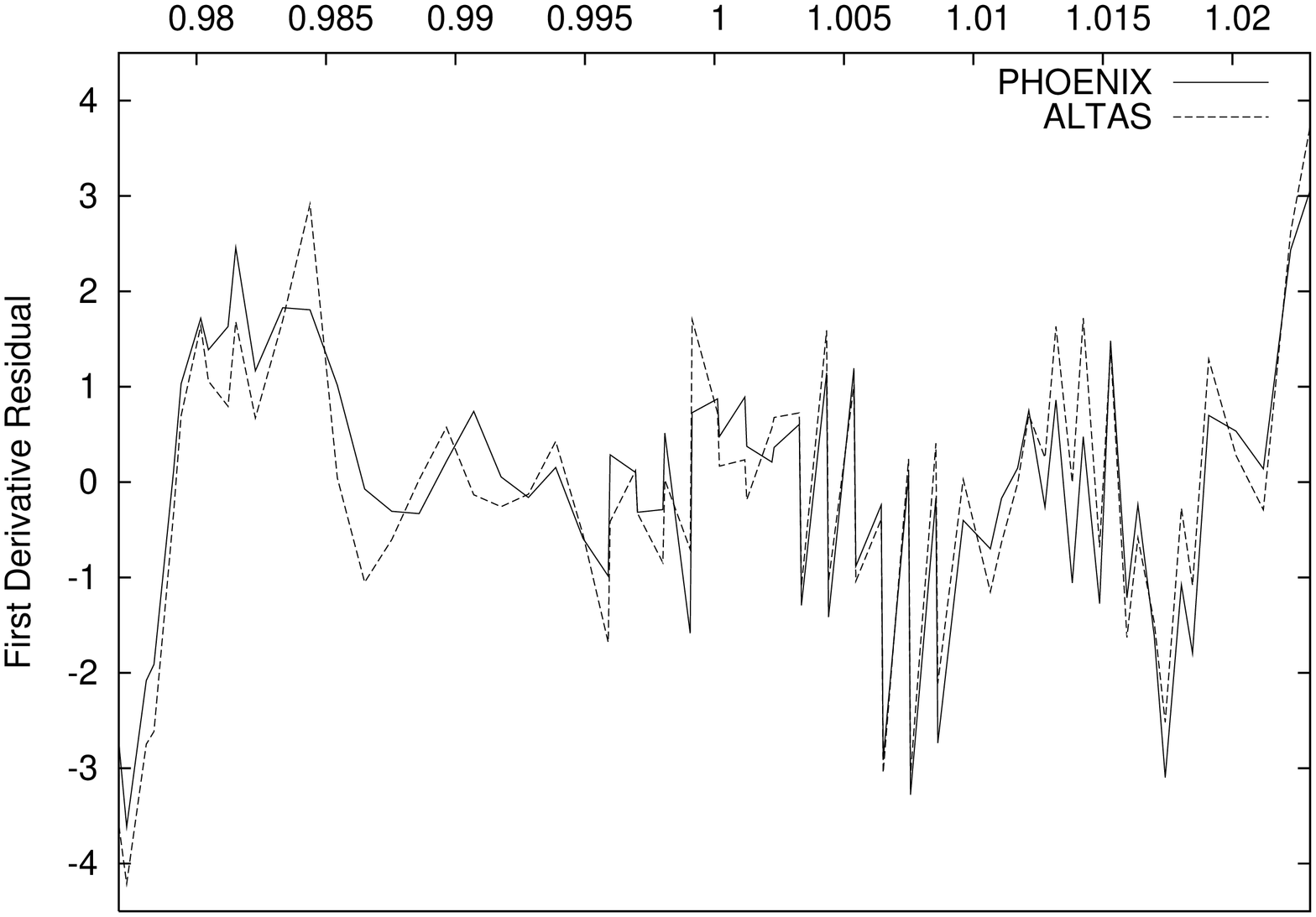,width=7.5cm,height=3.7cm,angle=270}
\caption{The first derivative profile centred at 4539\AA, plotted 
with the {\sc phoenix} and {\sc atlas} model atmospheres.  The phase
range is between second and third contact.  The lower plot shows the 
residual (observations - model) of this fit.}
\protect\label{chifit}
\end{figure}

\begin{figure}
\hspace{0.4cm}
\psfig{file=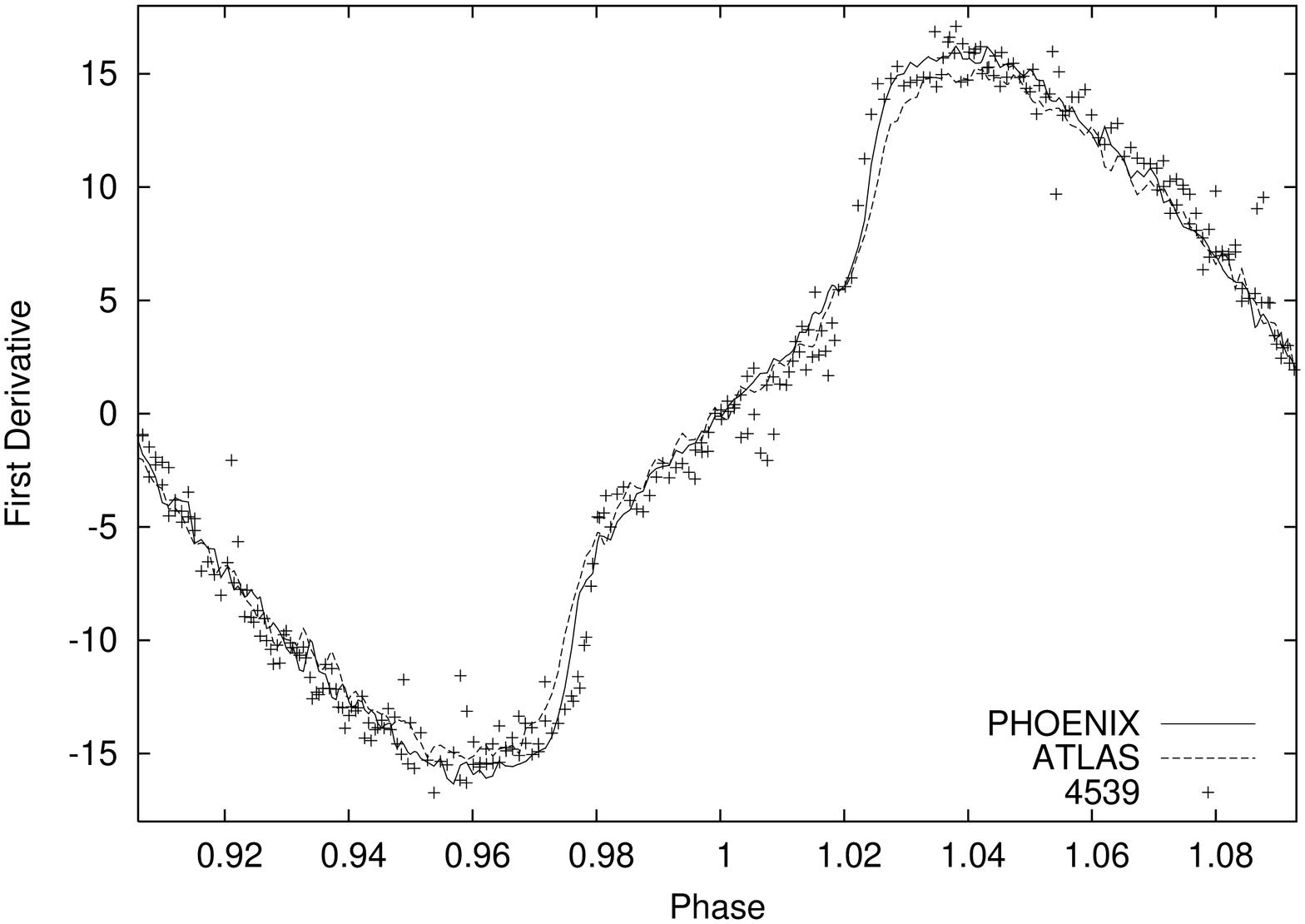,width=7.5cm,height=6cm,angle=270}
\caption{The first derivative profile centred at 4539\AA, plotted 
with the {\sc phoenix} and {\sc atlas} model atmospheres.  The phase
range is between first and fourth contact.}
\protect\label{chifitall}
\end{figure}

The fits to the observations are determined using (i) the totally
eclipsed section of the lightcurve between second and third contact
points only, and (ii) the region of the eclipse between the first and
fourth contact.  Figure~\ref{chifit}, and Figure~\ref{chifitall},
respectively show the results for case (i) and (ii) at the wavelength
4535\,\AA.  The $\chi^2$ values for all of the lightcurves are
summarised in Table~\ref{tchires}.

The results show that between first and fourth contact {\sc phoenix}
gives the best fit, except for the wavelength band centred at
3198\AA\ and 3707\AA.  However, the best fitting model in the region
of total eclipse between second and third contact shows a wavelength
dependence.  For wavebands centred at 4313\AA, 4539\AA, and 4743\AA\
{\sc atlas} provides the best fit, while {\sc phoenix} is the best 
fitting model atmosphere at shorter and longer wavelengths. 

\begin{table}
\fontsize{8}{12}\selectfont  
\begin{tabular}{c c c c c}


\hline
\hline

Wavelength & {\sc phoenix}(1) & {\sc atlas}(1)& {\sc phoenix}(2) &
{\sc atlas}(2) \\

\hline

3198  & 367.27 & 374.69 & 285.64 & 286.85 \\
3707 & 8.41 & 8.82 & 7.73 & 7.66 \\ 
4052 & 5.82 & 7.19 & 6.12 & 6.39 \\ 
4313 & 5.70 & 5.49 & 4.90 & 5.26 \\ 
4539 & 4.84 & 4.93 & 4.81 & 5.11 \\ 
4753 & 5.00 & 4.21 & 5.04 & 4.81 \\ 
4952 & 6.08 & 5.57 & 4.56 & 4.83 \\ 
5169 & 4.60 & 5.56 & 4.53 & 5.11 \\ 
5384 & 4.50 & 4.55 & 4.91 & 5.25 \\ 
5598 & 5.09 & 5.53 & 6.62 & 7.06 \\ 

\hline
\hline
\end{tabular}
\caption{Best fitting reduced $\chi^2$ values for lightcurves fitted in the
regions in the primary eclipse profile between (1)
first and fourth contact, and (2) second and third contact.  This
table is graphically shown in Figures~\ref{chi1t4}.}
\protect\label{tchires}
\end{table}

\begin{figure}
\psfig{file=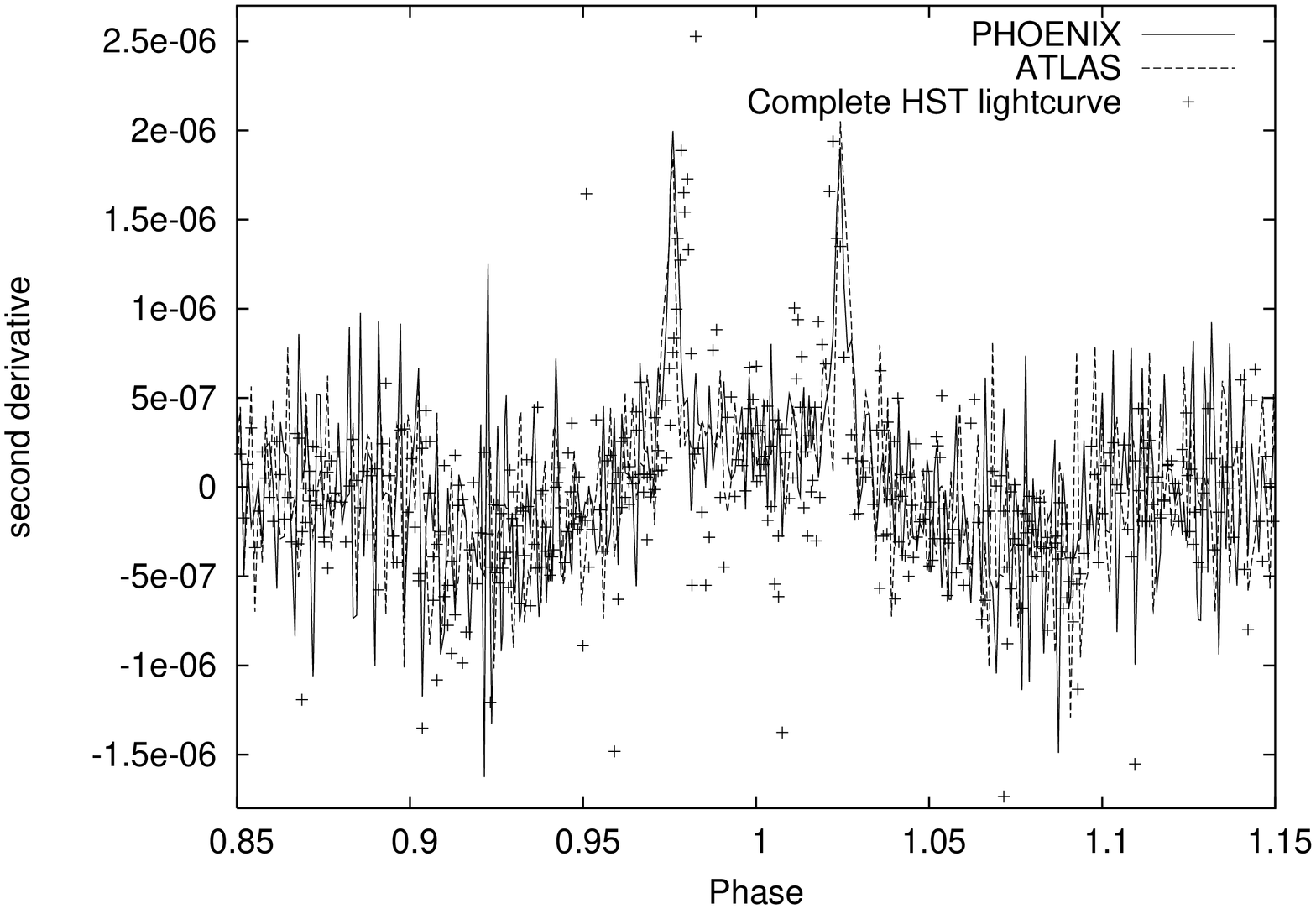,width=8.5cm,height=6cm,angle=270}
\caption{The second derivative profile using the complete HST
lightcurve, plotted with the {\sc phoenix} and {\sc atlas} model
atmospheres.}
\protect\label{sderiv}
\end{figure}

\begin{figure}
\psfig{file=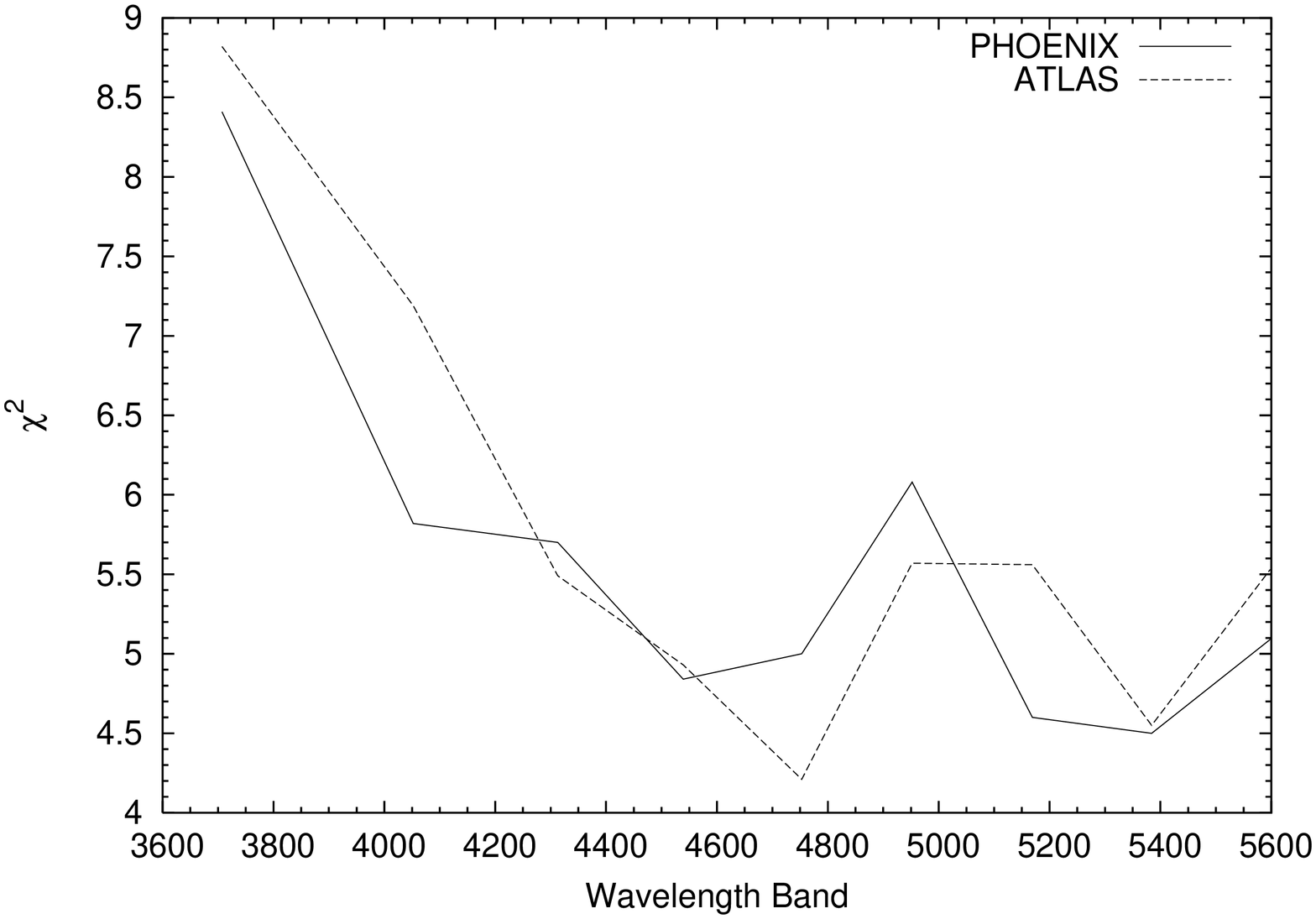,width=7.5cm,height=4cm,angle=270}
\psfig{file=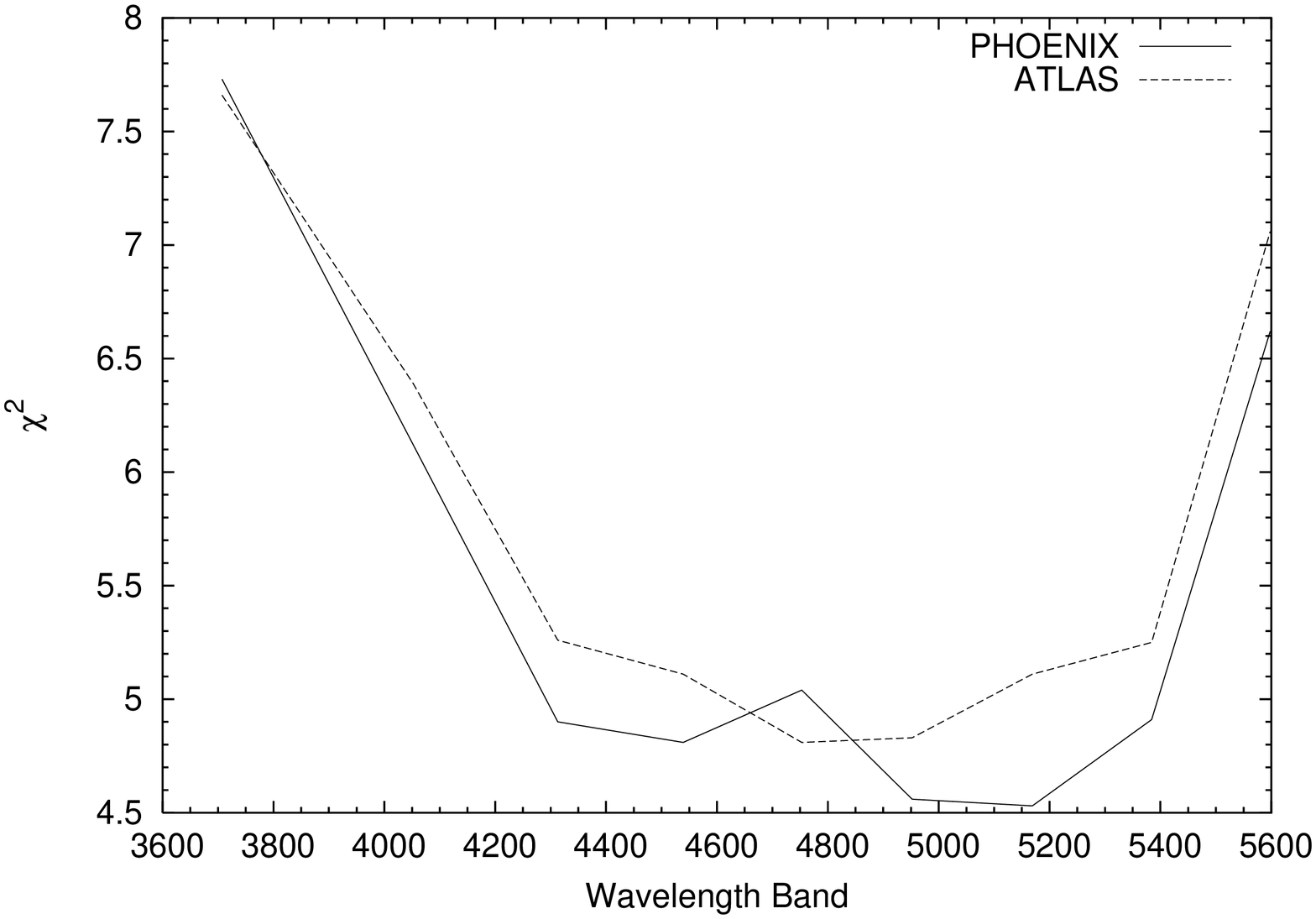,width=7.5cm,height=4cm,angle=270}
\caption{The variation of reduced $\chi^2$ as a function of wavelength for 
phases between second and third contact (top) and first and fourth 
contact (bottom) as tabulated in Table~\ref{tchires}.}
\protect\label{chi1t4}
\end{figure}

\section{Second derivative lightcurve fitting}

As with taking the first derivative of the lightcurve to determine the
rate of change of the slope in the eclipse profile, we now determine
the rate of change of the first derivative profile, i.e. the second
derivative of the eclipse profile.  The signal-to-noise ratio of the
observed data set was insufficient to determine useful second
derivative profiles for each of the 10 HST lightcurves.  Instead we
take the second derivative of the complete HST lightcurve and fit {\sc
phoenix} and {\sc atlas} model atmospheres centred at 4670\AA.  We fit
the second derivative profile between first and fourth contact as
between second and third contact we would only fit numerical noise.

The best fitting model atmosphere is {\sc phoenix} with a relative
$\chi^2$ of 7.02, while the {\sc atlas} model atmosphere has a
relative $\chi^2$ of 8.73.  The fit to the second derivative profile
for the {\sc phoenix} and the {\sc atlas} model atmospheres are shown
in Figure~\ref{sderiv}.  The jitter in the models is caused by 
numerical noise arising from the finite number of planar surface
elements used to model the star in the synthesis code.

\section{Discussion and Conclusions}

The best-fitting geometric parameters determined using {\sc phoenix}
and {\sc atlas} model atmospheres are in good agreement.  As shown in
Figure~\ref{centintwav}, the predicted specific intensity for the {\sc
atlas} models is greater at the limb than for {\sc phoenix} models,
which will consequently make the primary star bigger.  To compensate
for this the best-fitting temperature of the primary star is slightly
cooler than fitted using {\sc phoenix} models.  In this analysis we
solved the geometric system parameters of the lightcurve using
complete lightcurve rather than solving the parameters individually
for each of the 10 sub-lightcurves.  From Figure~\ref{centintwav}, we
would expect the best-fitting radii, solved using {\sc phoenix} and
{\sc atlas}, to be closer at redder wavelengths than at bluer
wavelengths, but would not differ enough to alter the conclusions of
this analysis.

We have shown the wavelength dependence of limb darkening by
sub-dividing the HST lightcurve into 10 bands of equal flux.  The
variation of flux between first and fourth contact shows that the limb
darkening decreases towards longer wavelengths, confirming published
limb darkening values, for example by \citet{claret00ldc4}, and as
observed on the Sun \cite{neckel94} and interferometrically in
K-giants by \cite{mozurkewich03}.  The splitting of the HST lightcurve
into 10 wavelength bands also highlights the presence of a time
variable bright bright feature, possibly an active region or plage on
the surface of the primary star, visible in the bluest wavelength band
(3198\AA).  The temporal variation of the bright feature is of the
order of 3 days as it comes into view on the primary star between the
second and third HST visit.

The ratio of the temperatures of the two stars has the effect that the
secondary star acts as a dark occulting disc that scans the surface of
the primary star.  During partial eclipse (i.e. between first and
second, and third and fourth contacts) the curvature of the lightcurve
provides information about the variation of the specific intensity
with limb angle of the primary star.  During total eclipse the
curvature of the lightcurve provides information on the variation of
specific intensity as a function of limb angle.  The first derivative
profile for each of the 10 HST wavelength bands clearly indicates the
change in slope as a function of phase.  Figure~\ref{fd2n10} shows the
wavelength variation of the gradient of the first derivative profile
of the HST lightcurves centred at 3707\AA\ and 5596\AA.  The slope of
the shorter wavelength is steeper than for the longer wavelength,
consistent with the limb darkening decreasing towards longer
wavelengths.

The best fitting model atmosphere is determined using $\chi^2$
comparison fit.  The first derivative profile of the modelled
lightcurves, generated with {\sc phoenix} and {\sc atlas} model
atmospheres, was fitted to the HST lightcurves.  The
results show that the majority of the results differ by less than
1$\sigma$ making the differences between the two models largely
insignificant.


Surface brightness reconstruction techniques such as Doppler imaging,
and eclipse mapping rely on the information content of surface areas
with different distances from the rotation axis (see review by
\citet{camerondoppler01}).  To detect starspots Doppler imaging uses 
the relative intensity contributions, calculated from model
atmospheres, to represent the different surface elements.  To
reconstruct an accurate surface brightness distribution it is
essential to know how parameters, such as the limb darkening, can
alter the intensity values across the stellar disc.  To date there are
many surface brightness images reconstructed using Doppler
imaging and eclipse mapping techniques on stars with similar spectral
types to SV Cam.  Examples include: He699, G2V \citep{jeffers02},
AE Phe G0V+F8V \citep{barnes04aephe}, Lu Lup, G2V
\citep{donati00rxj1508} and R58, G2V
\citep{marsden05}, reconstructed using {\sc 
atlas} plane-parallel model atmospheres.  The lightcurve modelling
results of this work clearly show that there is no distinguishable
difference between the two models using the high signal-to-noise SV
Cam observations.


The first derivative profiles show a small excess in the observed flux
at phase $\approx$ 0.9825 (just after second contact), compared with
the fitted model atmospheres.  In contrast, there is a slight decrease
in the observed flux at phase 1.015, i.e. just before third contact.
The increase and decrease of the first derivative at these points
could be evidence for an additional emission.  As this light excess is
located just before the second contact point, and the reverse just
before the third contact point, it could indicate that the very edge
of the secondary's limb is transparent to the light of the primary
star.

The signal-to-noise ratio of this data set was not high enough to
determine useful second derivative profiles for each of the 10 HST
lightcurves.  Instead we take the second derivative of the complete
HST lightcurve and fit {\sc phoenix} and {\sc atlas} model atmospheres
centred at 4670\AA\ to the observed lightcurve.  The jitter in the
models is caused by numerical noise arising from the finite number of
planar surface elements used to model the star in the synthesis code.
We fit the lightcurve between first and fourth contact as there is
insufficient structure to fit between second and third contact.  The
results show that the {\sc phoenix} model atmosphere code gives a
marginally better fit at the limb of the star.  However, {\sc phoenix}
does not provide an exact fit, which could indicate that the observed
cut-off in the limb intensity is steeper than predicted.  This could
explain why \citet{jeffersem05} could not completely remove the strong
discontinuities in the observed minus computed residual.

\section*{Acknowledgments}

The authors would like to thank J.R.Barnes for useful discussions.
SVJ acknowledges support from a PPARC research studentship and a
scholarship from the University of St Andrews while at St Andrews
University, and currently acknowledges support at OMP from a personal
Marie Curie Intra-European Fellowship within the 6$^{th}$ European
Community Framework Programme.

JPA was funded in part by a Harvard-Smithsonian CfA Postdoctoral
Fellowship and in part under contract with the Jet Propulsion
Laboratory (JPL) funded by NASA through the Michelson Fellowship
Program. JPL is managed for NASA by the California Institute of
Technology.

\bibliographystyle{mn2e}
\bibliography{iau_journals,master,ownrefs}

\appendix

\section[]{Determination of the percentage of spot coverage and polar
cap temperature for ATLAS model atmospheres}

In this appendix we describe the method used to determine the reduced
photospheric temperature due to the presence of many unresolvable
spots and the polar cap size, as quoted in Table~\ref{t-param} for the
ATLAS model atmospheres.  In a related paper, \citet{jeffersem05}
showed that in order to fit the HST lightcurve of SV Cam it was
necessary to reduce the photospheric temperature, to mimic the
peppering of the primary star's surface with small starspots, and to
include a polar cap.  In this appendix we determine the best-fitting
lightcurve solution to the SV Cam lightcurve using {\sc atlas} model
atmosphere.  It is important to do this to not to introduce an
inherent bias to the results of this paper.

\subsection{Unresolved spot coverage}

In this section determine the unresolved spot coverage, following the
method of \cite{jefferspc05}.  We assume that the unresolved spot
coverage comprises many small spots peppering the primary star, and a
polar cap which is not possible to resolve from a photometric
lightcurve.

\subsubsection{Temperature Fitting}

\begin{figure}
\psfig{file=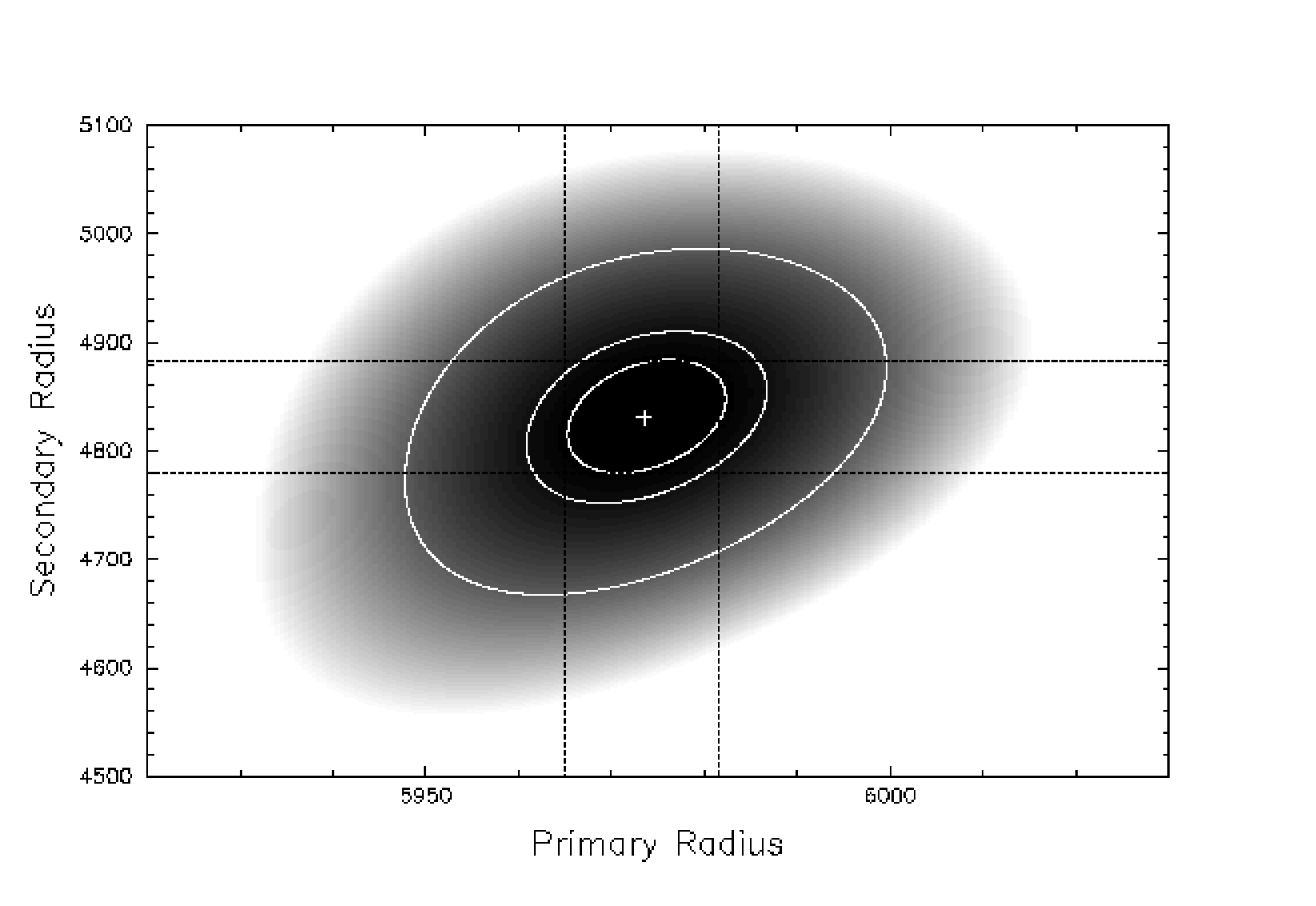,width=9.2cm,height=6.4cm,angle=270}
\caption{Contour plot of the $\chi^2$ landscape 
of the combined primary and secondary stars.  The minimum $\chi^2$
value occurs at 5973\,$\pm$\,31\,K, and 4831\,$\pm$\,103\,K, for the
primary and secondary stars respectively.  From the centre of the plot
the first contour ellipse represents the 1 parameter 1\,$\sigma$
confidence limit at 63.8\%, the second ellipse represents the 2
parameter 1\,$\sigma$ confidence limit at 63.8\% whilst the third
ellipse represents the 2 parameter 2.6\,$\sigma$ 99\% confidence
limit.}
\protect\label{cont_at}
\end{figure}

In the method of \cite{jefferspc05} the combination of the SV Cam
lightcurve and the Hipparcos parallax is used to determine the primary
and secondary temperatures.  Knowing the radii of the two stars we can
evaluate the flux contribution from the secondary star relative to
that of the primary star.  The best-fitting combination of primary and
secondary temperatures is determined using $\chi^2$ minimisation,
where a scaling factor $\gamma$ is included to ensure that the shape
of the spectrum is fitted rather than the absolute flux levels.  The
resulting $\chi^2$ landscape plot is shown in Figure~\ref{cont_at}.
The minimum value occurs at 5973$\pm$31\,K and 4831$\pm$103\,K for the
primary and secondary stars respectively.

\begin{figure}
\psfig{file=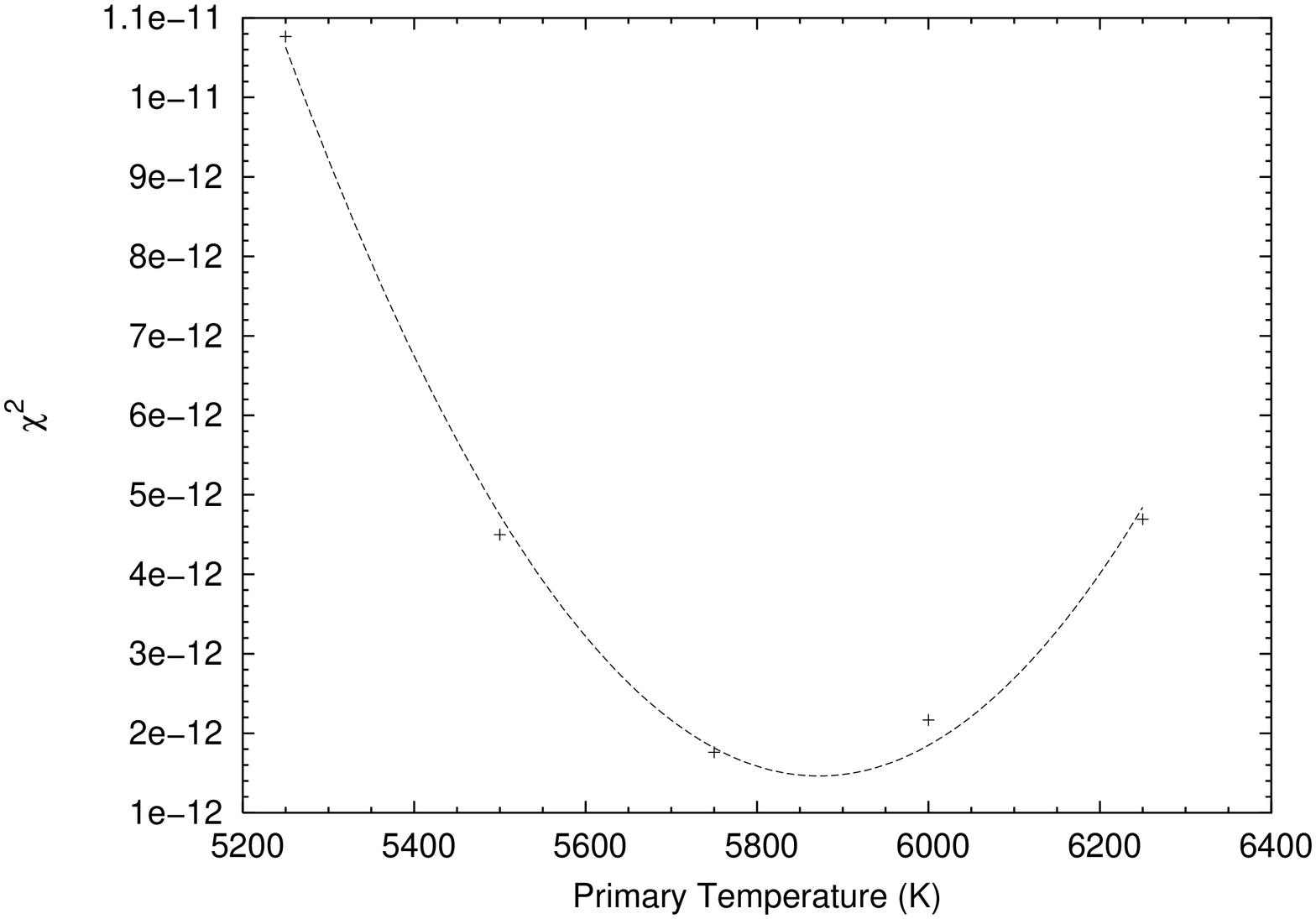,width=9.1cm,height=6.3cm,angle=270}
\caption{Parabolic fit to determine the primary temperature minimum 
to be 5872$\pm$53\,K}
\protect\label{pri_temp_atls}
\end{figure}

The temperature of the primary star is then determined by isolating
the primary star's spectrum.  To achieve this we subtracted a spectrum
outside of the primary eclipse from one inside of the eclipse which
results in the spectrum of the primary star but with the radius of the
secondary star.  The temperature is fitted using $\chi^2$ minimisation,
where the minimum temperature is determined by a parabolic fit.  This
results in a minimum primary temperature of 5872$\pm$59\,K
(Figure~\ref{pri_temp_atls}), with the errors determined by setting
$\Delta\chi^2$=1.

\subsubsection{Fractional Starspot Coverage}

\begin{figure}
\psfig{file=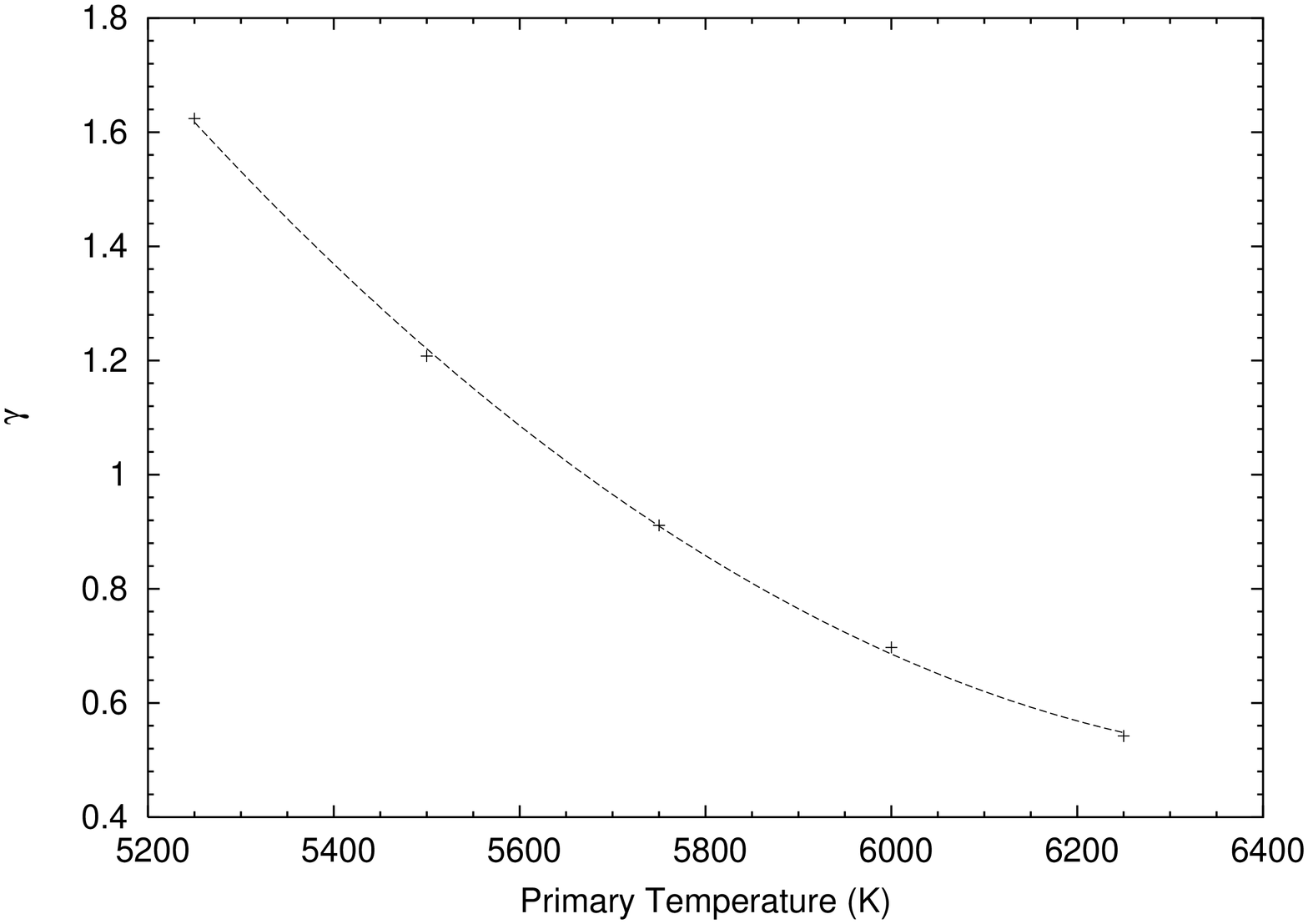,width=9.1cm,height=6.3cm,angle=270}
\caption{Scaling factor ($\gamma$) as a function of Primary
Temperature ({\sc atlas} models).  The scaling 
factor for T=5872\,K is determined by a quadratic fit to these
points.}
\protect\label{pri_scal_atls}
\end{figure}

Following the conclusions of \cite{jefferspc05} we attribute the
missing flux to be indicative of small unresolvable spots on the
primary star's surface.  The dark starspot filling factor is given by:

\begin{equation}
\alpha = 1 - \gamma
\label{e-alpha}
\end{equation}

where $\alpha$ is the fractional starspot coverage and $\gamma$ is the
scaling factor.  The interpolated scaling factor for the primary
temperature is determined as shown in Figure~\ref{pri_scal_atls}.  The
scaling factor is 0.76, resulting in a fractional coverage of dark
starspots of 24\%.

\subsubsection{Polar Cap}

\begin{figure}
\psfig{file=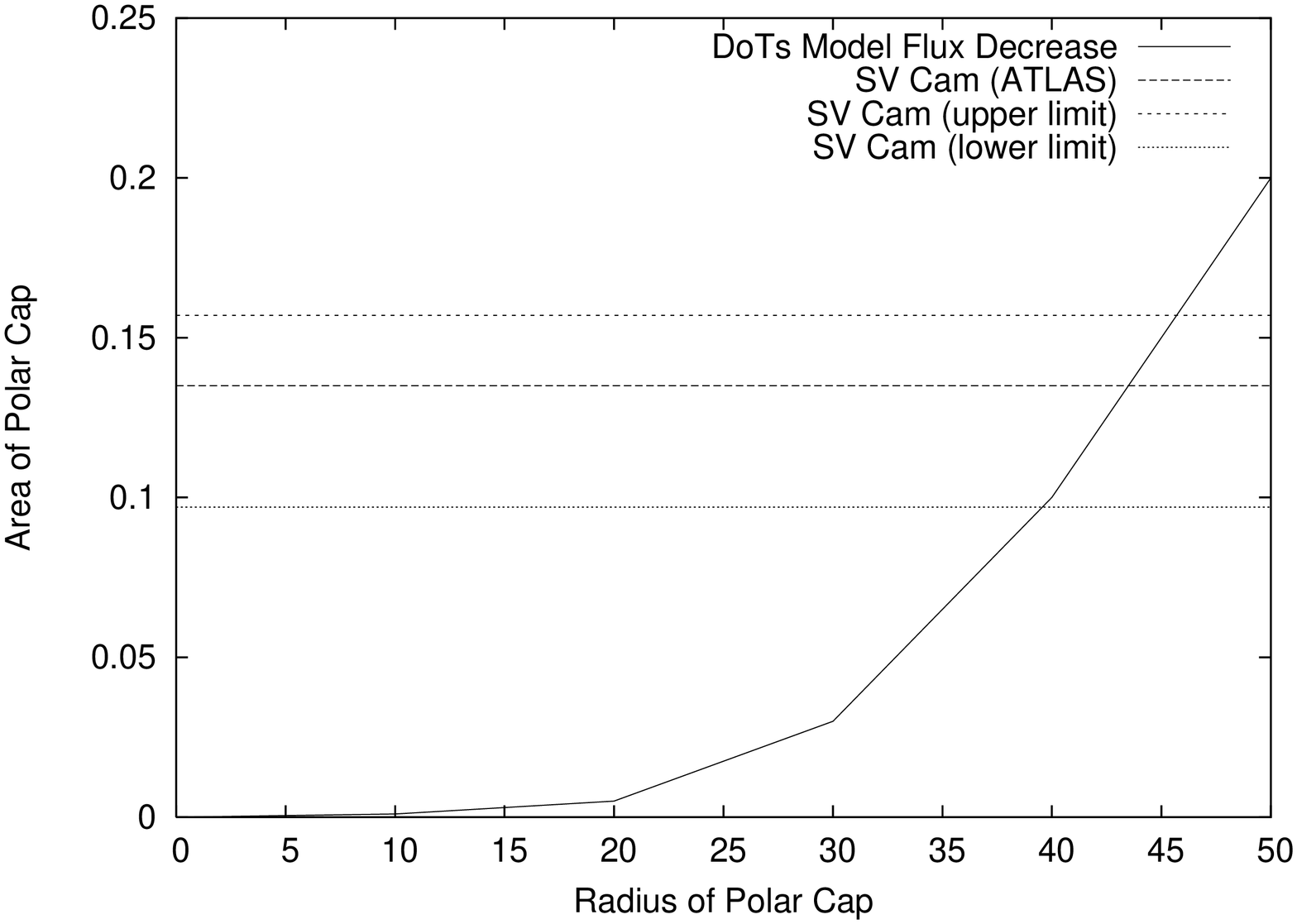,width=9.1cm,height=6.3cm,angle=270}
\caption{Fractional decrease in the stellar flux of SV Cam as a
function of theoretical polar cap size}
\protect\label{pcap_atls}
\end{figure}

The determination of the spot coverage fraction only accounts for the
flux deficit in the eclipsed equatorial latitudes of the primary
star.  Extending the 24\% spot coverage to the entire surface of the
primary star (as described by \citep{jefferspc05}) we find that there
is an additional 13.5\% flux deficit.  The binary eclipse-mapping code
DoTs is used to model artificial polar spots on the surface of SV Cam,
to include effects resulting from the star being a sphere and not a
disc, limb and gravity darkening and spherical oblateness.  We model
the fractional decrease in stellar flux as a function of polar spot
size and determine the polar spot radius to be 43.5$\pm$6$^\circ$
(Fig. ~\ref{pcap_atls}).
 
\subsection{Lightcurve Modelling}

To compare ATLAS and PHOENIX model atmospheres we need to determine
the best-fitting binary system parameters to the observed SV Cam
lightcurve for each model atmosphere separately to avoid introducing a
an inherent bias to our results.  We include the presence of high
unresolvable spot coverage and polar caps in the lightcurve fit by
using the method of \cite{jeffersem05}.  Such spot coverage has
been shown by that method to have a significant impact on the binary
system parameters.

\subsubsection{Reduced Photospheric Temperature}

\begin{figure}
\psfig{file=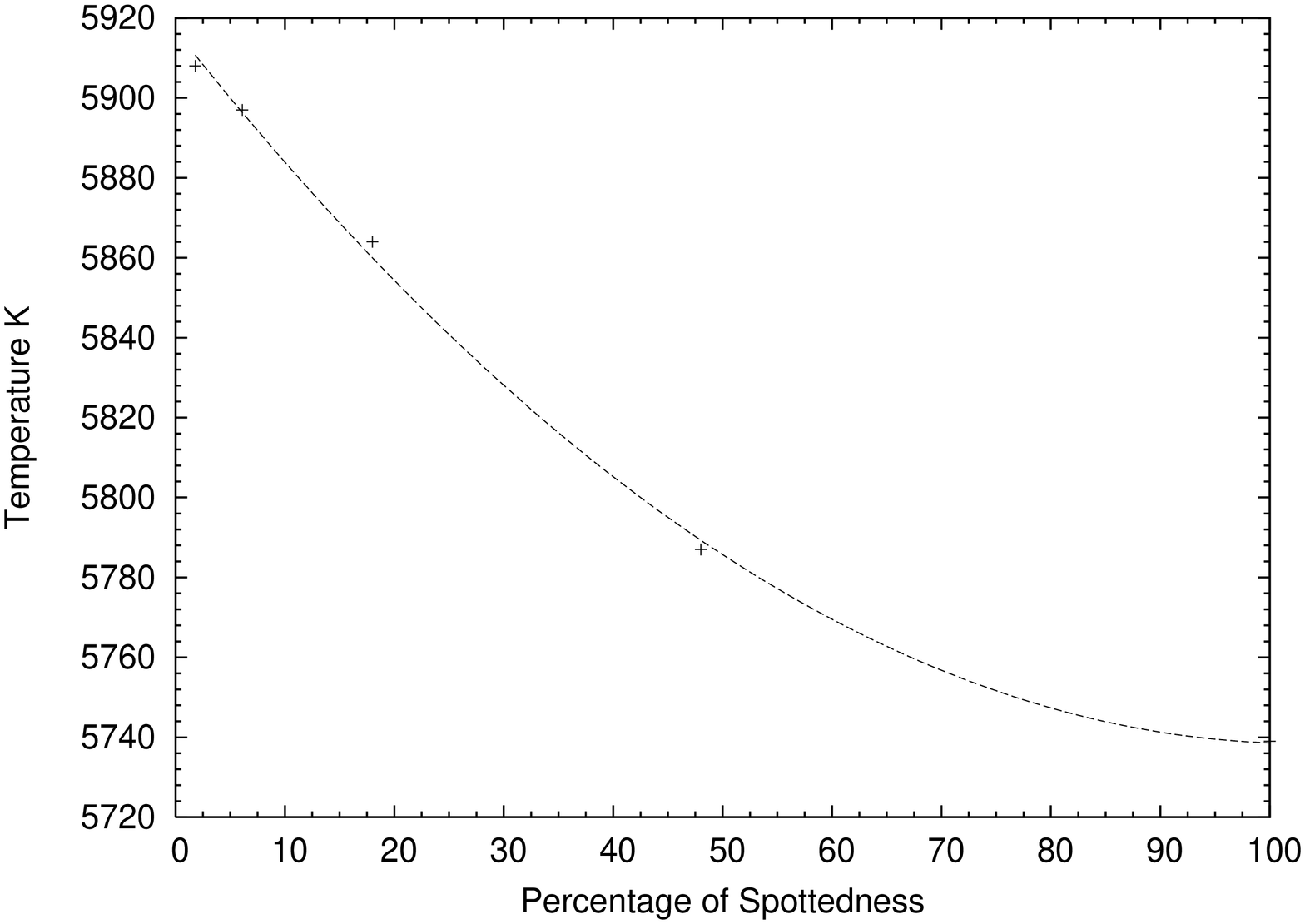,width=9.1cm,height=6.3cm,angle=270}
\caption{The decrease of the primary star's apparent photospheric
temperature as a function of the percentage of spot coverage on its 
surface.}
\protect\label{spotredats}
\end{figure}

The peppering of small starspots poses a limitation on image
reconstruction techniques from photometric lightcurves such as Maximum
Entropy eclipse mapping \citep{jeffersfs05}.  To model the presence of many
dark unresolvable starspots we decrease the apparent photospheric
temperature of the star.  To determine the reduction in the
photospheric temperature of the star we model starspot distributions
equating to 1.8\%, 6.1\%, 18\%, 48\% and 100\% of the stellar surface
on an immaculate SV Cam.  For each model the initial photospheric
temperature is 5904\,K and the spot temperature is 4400\,K.  Each of
these starspot distributions is modelled as a photometric lightcurve.
Using the Maximum Entropy eclipse mapping method we determine the
best-fitting temperature to each model lightcurve using a $\chi^2$
grid-search method.  A quadratic fit to the best-fitting temperatures
show that for a starspot coverage of 24\%, the apparent photospheric
temperature is 5840\,K (Figure~\ref{spotredats}).

\subsubsection{Polar Cap}

\begin{figure}
\psfig{file=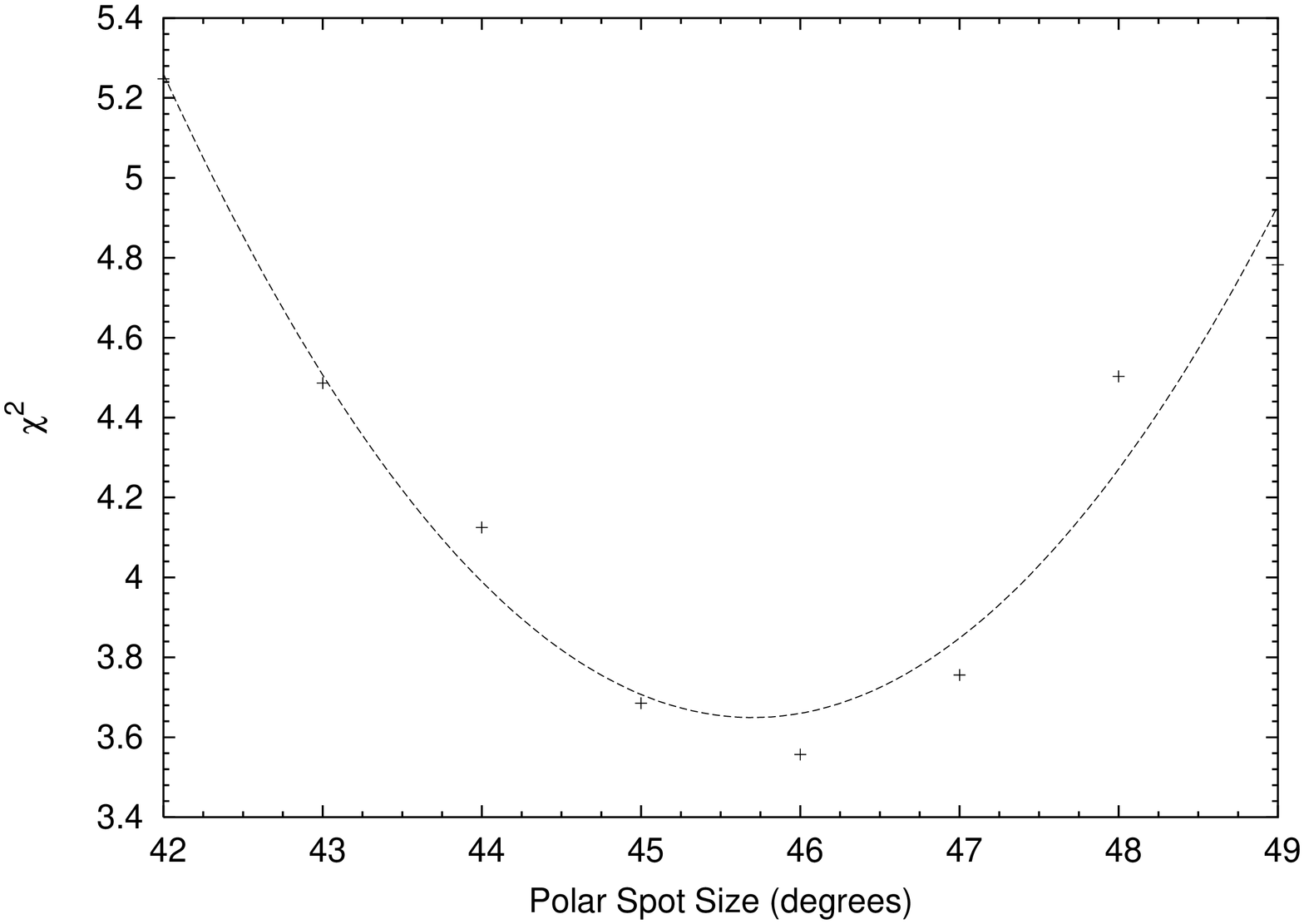,width=8.7cm,height=5.6cm,angle=270}
\caption{Quadratic fit to the variation of $\chi^2$ as a function of 
polar spot size.}
\label{chipcap}
\end{figure}

\begin{figure}
\psfig{file=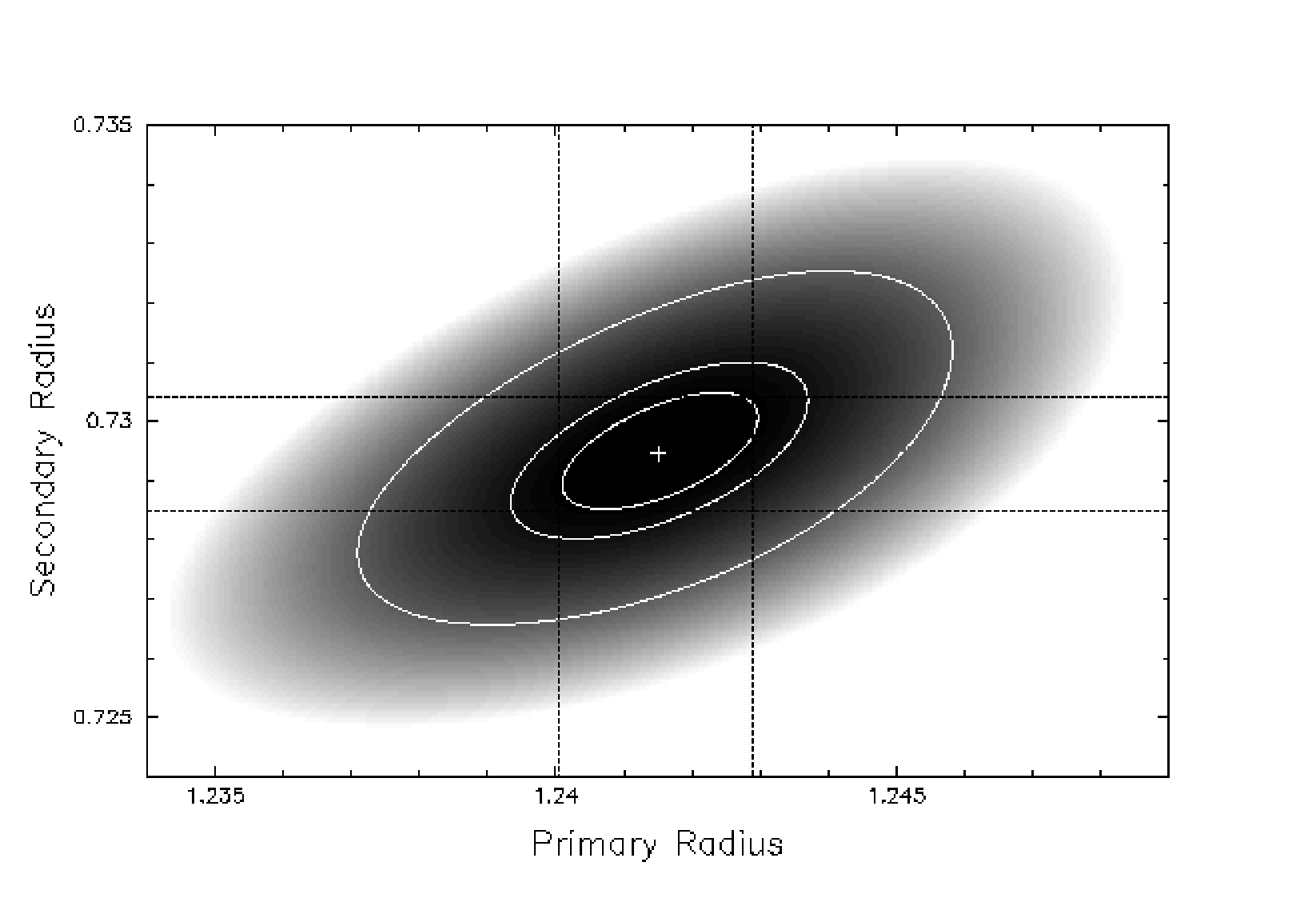,width=8.7cm,height=6.2cm,angle=270}
\caption{Contour plot of the $\chi^2$ landscape for the primary and
secondary radii with a 45.7$^\circ$ polar spot.  From the centre of
the plot, the first contour ellipse is the 1 parameter 1$\sigma$
confidence limit at 63.8\%, the second contour ellipse is the 2
parameter 1$\sigma$ confidence limit at 63.8\%, and the third contour
ellipse is the 2 parameter 2.6$\sigma$ confidence limit at 99\%.}
\label{pcaprad}
\end{figure}

We include a polar cap in our analysis to verify the results in the
previous section, following the method of \cite{jeffersem05}.  The
polar spot is assumed to be at 4500\,K, circular, centred at the
pole, and is in addition to the peppered spot distribution as
described above.  For each polar spot size (from 40$^\circ$ to
50$^\circ$) the minimum primary and secondary radii are determined
using a $\chi^2$ contour map.  These minimum $\chi^2$ values are
plotted as a function of polar spot size in Figure~\ref{chipcap}.  The
best fitting polar cap size, 45.7$^\circ$ was determined from the
minimum of a quadratic function fitted to these points.  The grid
search of radii is repeated using a fixed value for the polar spot
size.  The results for the $\chi^2$ minimisation are summarised in
Table~\ref{t-param} shown as a contour map in Figure~\ref{pcaprad}.

\subsection{Summary}

We have shown in this section that the best fitting binary system
parameters determined using the {\sc atlas} model atmosphere are in
agreement with those determined using {\sc phoenix} model atmospheres
\citep{jefferspc05}.  This further shows that at there is no significant
difference between {\sc phoenix} and {\sc atlas} model atmospheres at
the F9V+K4V spectral types.  The results are summarised in
Table~\ref{t-param}.

\bsp
\protect\label{lastpage}

\end{document}